\documentclass[twocolumn,superscriptaddress,showpacs,preprintnumbers,nofootinbib,amsmath,amssymb,floatfix,aps]{revtex4-1}
\usepackage{dcolumn}% Align table columns on decimal point
\usepackage{bm}% bold math
\usepackage{color}

\usepackage{graphicx}% Include figure files
\usepackage{amsmath}
\usepackage{amssymb}

\def\lsim{\buildrel < \over {_{\sim}}}

\def\beq{\begin{equation}}
\def\eeq{\end{equation}}
\def\be{\begin{eqnarray}}
\def\ee{\end{eqnarray}}
\def\ra{\rangle}
\def\la{\langle}

\begin{document}
\title{Electromagnetic scaling functions within the Green's Function Monte Carlo approach}

\author{N. Rocco}
\affiliation{Instituto de F\'\i sica Corpuscular (IFIC), Centro Mixto CSIC-Universidad de Valencia, E-46071 Valencia, Spain} 
\author{L. Alvarez-Ruso}
\affiliation{Instituto de F\'\i sica Corpuscular (IFIC), Centro Mixto CSIC-Universidad de Valencia, E-46071 Valencia, Spain}
\author{A. Lovato}
\affiliation{Physics Division, Argonne National Laboratory, Argonne, Illinois 60439, USA}
\author{J. Nieves}
\affiliation{Instituto de F\'\i sica Corpuscular (IFIC), Centro Mixto CSIC-Universidad de Valencia, E-46071 Valencia, Spain} 

%%%%%%%%%%%%%%%%%%%%%%%%%%%%%%%%%%%%%%%%%%%%%%%%%%%%%%%%%%%%%%%%%%%%%%%%%
\date{\today}
%%%%%%%%%%%%%%%%%%%%%%%%%%%%%%%%%%%%%%%%%%%%%%%%%%%%%%%%%%%%%%%%%%%%%%%%%
\begin{abstract}
We have studied the scaling properties of the electromagnetic response functions of $^4$He and $^{12}$C nuclei computed by the Green's Function Monte Carlo approach, retaining only the one-body current contribution. Longitudinal and transverse scaling functions have been obtained in the relativistic and non relativistic cases and compared to experiment for various kinematics. The characteristic asymmetric shape of the scaling function exhibited by data emerges in the calculations in spite of the non relativistic nature of the model. The results are consistent with scaling of zeroth, first and second kinds. Our analysis reveals a direct correspondence between the scaling and the nucleon-density response functions.   
\end{abstract}
\maketitle
%%%%%%%%%%%%%%%%%%%%%%%%%%%%%%%%%%%%%%%%%%%%%%%%%%%%%%%%%%%%%%%%%%%%%%%%%

\section{Introduction}
A realistic description of the electromagnetic response of atomic nuclei is a challenging many-body problem as it requires an accurate understanding of both the nuclear dynamics and of the interaction vertex.
In this regard a valuable strategy consists in analyzing the scaling properties of nuclear response functions in a variety of kinematic setups \cite{West:1974ua,Day:1987az,Donnelly:1998xg}.  Scaling of the first kind is said to occur when the electron-nucleus cross section or longitudinal/transverse response functions, divided by an appropriate function describing the single-nucleon physics, do no longer depend on two variables (for example energy transfer $\omega$ and absolute value of the 3-momentum transfer ${|\bf q|}$  in the Laboratory frame),
but only upon a specific function of them, which defines the scaling variable. Scaling of the second kind takes place when there is no dependence on the nuclear species. Finally, the simultaneous occurrence of both kinds of scaling is denoted as superscaling~\cite{Alberico:1988bv}. 

Superscaling is exactly fulfilled by the Global Relativistic Fermi gas (GRFG) model, for which a simple and symmetric scaling function can be derived in terms of the dimensionless scaling variable $\psi$~\cite{Barbaro:1998gu} (explicit expressions are provided in Sec.~\ref{GRFG} below). However, contrary to the GRFG model predictions, the results extracted from experimental data reveal an asymmetric shape of the scaling function, with a tail that extends to high values of $\psi$ (and $\omega$)~\cite{Amaro:2004bs}. These results represent a strong constraint for theoretical models of electron scattering reactions. Extensive studies with a large variety of models reveal the importance of a proper description of the interaction of knocked-out nucleons with the residual nucleus---final state interactions (FSI)---to obtain the tail of the scaling function~\cite{Caballero:2005sj,Caballero:2006wi,Caballero:2007tz,Meucci:2009nm,Antonov:2011bi}. The authors of Refs.~\cite{Caballero:2005sj,Caballero:2006wi} argue that, while this asymmetry in the scaling function is largely absent in non-relativistic mean-field models, it can be recovered within the relativistic impulse approximation, given that FSI are described using a strong relativistic mean field (RMF) potential. Asymmetric scaling functions also emerge in semi-relativistic models when FSI are described by local potentials derived from the RMF one~\cite{Caballero:2007tz}. On the other hand, the comparison between semi-relativistic and relativistic results shows a breakdown of the zeroth-kind scaling, {i.e.} different scaling functions in the longitudinal and transverse channel, only when the fully relativistic mean field approach is employed. According to Ref.~\cite{Caballero:2007tz} this effect has been ascribed to the dynamical enhancement of the lower component of the Dirac spinors, which are not present in the semi-relativistic approach.

In this work we analyze the scaling properties exhibited by Green's Function Monte Carlo (GFMC).  GFMC is an {\em ab initio} method allowing for a very accurate description of the properties of $A\leq 12$ nuclei, in which the dynamics of constituent nucleons are fully considered~\cite{Lovato:2013cua,Lovato:2014eva,Lovato:2015qka}. The longitudinal and transverse electromagnetic response functions of $^{12}$C, recently computed within GFMC turn out to be in very good agreement with experiment, when two-body currents are accounted for~\cite{Lovato:2016gkq}. Despite this remarkable result, GFMC is currently limited to $^{12}$C because of the exponentially growing cost of the calculation with the number of nucleons. In addition to that, the inclusion of relativistic kinematic and baryon resonance production would involve non trivial difficulties.

The study of the behavior of the scaling functions obtained from the GFMC calculations, while being interesting in its own right, is aimed at elucidating the role of initial and final state correlations in the asymmetric shape of the scaling function.

In Section~\ref{scaling} we review the derivation of the electron-nucleus cross section, as well as its expression  in terms of longitudinal and transverse response functions, which are  necessary to introduce the concept of scaling.  In Section~\ref{GFMC}, the main elements of the Green's Function Monte Carlo approach are briefly outlined, while in Section~\ref{GRFG}  we explicitly derive the expression of the longitudinal and transverse scaling functions in the context of the GRFG model, both in the relativistic and non relativistic cases. In Section~\ref{results} we report the results of our analysis of the scaling features of the GFMC response functions for $^4$He and $^{12}$C nuclei and in different kinematics. We then discuss a novel interpretation of the longitudinal and transverse scaling function in terms of the nucleon-density function. Finally, in Section \ref{conclusion} we summarize our findings and state the conclusions.

\section{Scaling of the nuclear electromagnetic response within the Green's Function Monte Carlo approach}
\label{scaling}

In the one-photon-exchange approximation, the  double differential electron-nucleus cross section can be written
in the form
\beq
\label{xsec}
\frac{d^2\sigma}{d E_{e^\prime} d\Omega_{e^\prime}}=\frac{\alpha^2}{q^4}\frac{E_{e^\prime}}{E_e}L_{\mu\nu}W^{\mu\nu} \ ,
\eeq
where $k_e=(E_e,{\bf k}_e)$ and $k_{e^\prime}=(E_{e^\prime},{\bf k}_{e^\prime})$ are the laboratory four-momenta of the incoming and outgoing electrons, respectively; $\alpha \simeq 1/137$ is the fine structure constant, $d\Omega_{e^\prime}$, the differential solid angle in the direction of ${\bf k}_{e^\prime}$, and $q=k_e - k_{e^\prime} =(\omega,{\bf q})$  the four momentum transfer.  
The leptonic tensor is given by
\begin{align}
  L^{\mu\nu}=2 \left( k_{e^\prime}^\mu k_e^\nu+ k_e^\mu k_{e^\prime}^\nu- g^{\mu\nu}k_{e^\prime}\cdot k_e \right)\,.  
\end{align}
The hadronic tensor encompasses  the electromagnetic transitions from the target nucleus to all possible final states. It is thus given by
\begin{align}
\label{response:tensor}
W^{\mu \nu} =\sum_f \langle  0| {J^\mu}^\dagger(q) | f \rangle  \langle  f | J^\nu(q) |   0 \rangle
\, \delta^{(4)}(P_0+q-P_f)   \ ,
\end{align}
where $| 0 \rangle$ and $| f \rangle$ denote the initial and final hadronic states with four-momenta $P_0 = ( E_0,{\bf p}_0 )$ and $P_f = (E_f,{\bf p}_f) $, while $J(q)$ is the electromagnetic nuclear current operator.

Equation \eqref{xsec} can be rewritten in terms of two response functions, denoted by $R_L({\bf q}, \omega)$ 
and $R_T({\bf q},\omega)$, describing interactions with longitudinally (L) and transversely (T) polarized photons, 
respectively. The resulting expression reads 
\begin{align}
\frac{d^2\sigma}{d E_{e^\prime} d\Omega_{e^\prime}} &  =\left( \frac{d \sigma}{d\Omega_{e^\prime}} \right)_{\rm{M}} \Big[  A_L(|{\bf q}|,\omega,\theta_{e^\prime})  R_L(|{\bf q}|,\omega) \nonumber \\
& + A_T(|{\bf q}|,\omega,\theta_{e^\prime})  R_T(|{\bf q}|,\omega) \Big] \ ,
\end{align}
where 
\begin{align}
A_L = \Big( \frac{q^2}{{\bf q}^2}\Big)^2  \ \ \ , \ \ \ A_T = -\frac{1}{2}\frac{q^2}{{\bf q}^2}+\tan^2\frac{\theta_e}{2}  \ , 
\end{align}
and
\begin{align}
\label{Mott}
\left( \frac{d \sigma}{d \Omega_{e^\prime}} \right)_{\rm{M}}= \left[ \frac{\alpha \cos(\theta_{e^\prime}/2)}{2 E_{e^\prime}\sin^2(\theta_{e^\prime}/2) }\right]^2
\end{align}
is the Mott cross section. The L and T response functions can be readily expressed in terms of specific components of the hadronic tensor. Choosing the $z$-axis along the direction of the momentum transfer one finds
\begin{align}
\label{RL}
R_L & = W^{00}\ ,\\
\label{RT}
R_T &= \sum_{ij=1}^3\Big(\delta_{ij}-\frac{q_iq_j}{{\bf q}^2}\Big)W^{ij} \ . 
\end{align}

\subsection{The Green's function Monte Carlo approach}
\label{GFMC}

GFMC provides a suitable framework to carry out accurate calculations of a variety of nuclear properties 
in the non relativistic regime, typically corresponding to $|{\bf q}| \lsim 500 \ {\rm MeV}$ (for a recent review of Quantum Monte Carlo methods for nuclear physics see, e.g., Ref. \cite{Carlson:2014vla}). 

The longitudinal and transverse response function are given by
\begin{align}
R_L({\bf q}, \omega)&= \sum_f \la 0|\rho^\dagger({\bf q})|f\ra\la f|\rho({\bf q})|0\ra\delta(\omega+E_0-E_f)\,, \nonumber\\
R_T({\bf q}, \omega)&= \sum_f \la 0|{\bf j}_T^\dagger({\bf q})|f\ra\la f|{\bf j}_T({\bf q})|0\ra\delta(\omega+E_0-E_f) \,,
  \label{resp:GFMC}
 \end{align}
 where $\rho({\bf q})$ and ${\bf j}_T({\bf q})$ denote non-relativistic reductions of the nuclear-charge and transverse-current operators, respectively \cite{Carlson:2001mp}.
Valuable information on the L and T responses can be obtained from their Laplace transforms, also referred to as Euclidean responses
\beq
\widetilde{E}_{T,L}({\bf q}, \tau)= \int_{\omega_{\rm{el}}}^\infty \,{d\omega} e^{-\omega \tau}R_{T,L}({\bf q}, \omega)\ .
\eeq
The lower integration limit $\omega_{\rm{el}}= {\bf q}^2/2M_A$, $M_A$ being the mass of the target nucleus, is the elastic scattering threshold---corresponding to the 
$|f \rangle = |0 \rangle$ term in the sum of Eq. \eqref{response:tensor}---whose contribution is excluded.

Within GFMC, the Euclidean responses are evaluated from 
\begin{align}
\nonumber
\widetilde{E}_L({\bf q},\tau) & = \langle 0| \rho^\dagger({\bf q}) e^{-(H-E_0)\tau}  \rho({\bf q})|0\rangle \\ 
& -  |\langle 0 | \rho({\bf q}) | 0 \rangle|^2 e^{-\omega_{\rm el} \tau} \ ,
\label{eq:eucL_mat_el}
\end{align}
and 
\begin{align}
\nonumber
\widetilde{E}_T({\bf q},\tau) & = \langle 0| {\bf j}_T^\dagger({\bf q}) e^{-(H-E_0)\tau} {\bf j}_T({\bf q})|0\rangle \\ 
& -  |\langle 0 | {\bf j}_T({\bf q}) | 0 \rangle|^2 e^{-\omega_{\rm el} \tau} \ .
\label{eq:eucT_mat_el}
\end{align}
Note that, although the states $|f \rangle \neq | 0 \rangle$ do not appear explicitly in Eqs. \eqref{eq:eucL_mat_el} and \eqref{eq:eucT_mat_el}, the Euclidean responses  
include the FSI effects  of the particles involved in the electromagnetic interaction, both among themselves and with the spectator nucleons. 

The inversion of the Laplace transform, needed to retrieve the energy dependence of the responses, is long known to involve severe difficulties.   However, maximum-entropy techniques, based on Bayesian inference arguments, have been successfully exploited to perform accurate inversions, supplemented by reliable estimates of the theoretical uncertainty. In the case of $^{12}$C, particular care has to be devoted to the subtraction of contributions arising from elastic scattering and the transitions to the low-lying $2^+$, $0^+_2$, and $4^+$ states~\cite{Lovato:2015qka}.

\subsection{Scaling within the relativistic Fermi gas model}
\label{GRFG}

The easiest, albeit quite crude approximation, to describe the hadron tensor consists on using the GRFG model. Within this approach the scattering process is assumed to take place on a single nucleon with four-momentum $p=(E({\bf p}),{\bf p})$, where $E({\bf p})=\sqrt{|{\bf p}|^2+m^2}$, $m$ being the nucleon mass. The requirement that the struck nucleon is in the target nucleus implies that $|{\bf p}|$ is smaller than the Fermi momentum $p_F$. Furthermore, the outgoing nucleon with four-momentum ${p^\prime}^\mu=(p+q)^\mu$ should lay above the Fermi surface. The expression of the hadron tensor describing the response of the target nucleus then reads
\begin{align}
W^{\mu\nu}=& \frac{3 \mathcal{N} }{4\pi p_F^3}\int d^3p\frac{m^2}{E({\bf p})E({\bf p+q})}\ w^{\mu\nu}(p+q,p)\nonumber\\
 &\times \theta(p_F-|{\bf p}|)\theta(|{\bf p+q}|-p_F)\nonumber\\
 &\times \delta(\omega+E({\bf p})-E({\bf p+q}))\ .
 \label{had:tens:FG}
\end{align}
Once we only discuss symmetric nuclei, $\mathcal{N}$ denotes both the number of protons and neutrons in the nucleus.  The single-nucleon response tensor $w^{\mu\nu}(p+q,p)$ encodes the response of a system in which a nucleon with 4-momentum $p$ in the initial state is scattered by a (virtual) photon, leading to a final state with a nucleon carrying a 4-momentum $(p+q)$. The following general expression 
\begin{align}
\label{w12}  
w^{\mu\nu}(p+q,p)=& - W_1(\tau)\Big(g^{\mu\nu}-\frac{q^\mu q^\nu}{q^2}\Big)\nonumber\\
&+ W_2(\tau)\frac{1}{m^2}\Big(p^\mu-\frac{p\cdot q}{q^2}q^\mu\Big)\nonumber\\
&\times \Big( p^\nu -\frac{p\cdot q}{q^2}q^\nu\Big)\ ,
\end{align}
where $\tau= -q^2/4m^2= Q^2/4m^2\geq0$, holds. It is well known that the nucleon structure functions $W_{1,2}$ can be written in terms of the proton and neutron electric and magnetic form factors as
\begin{align}
W_1(\tau)=& \tau G^2_M(\tau)\ ,\nonumber\\
W_2(\tau)=& \frac{G_E^2(\tau)+\tau G_M^2(\tau)}{(1+\tau)}\ ,
\end{align}
and
\begin{align}
G_E(\tau)&= G^p_{E}(\tau)\frac{1}{2}(1+\tau_{z,i})+ G_E^n(\tau)\frac{1}{2}(1-\tau_{z,i})\ ,\nonumber\\
G_M(\tau) &= G_M^p(\tau)\frac{1}{2}(1+\tau_{z,i})+ G_M^n(\tau)\frac{1}{2}(1-\tau_{z,i})\ ,
\end{align}
where $\tau_{z,p/n} = \pm 1$. 

Using the GRFG model to parametrize the nuclear amplitudes, the integral entering Eq.~\eqref{had:tens:FG} can be analytically solved. We start by evaluating the function
\begin{align}
F(p_F,q)= & \frac{3 \mathcal{N} }{4\pi p_F^3}\int d^3p\ \mathcal{F}(p_F,q,{\bf p})\ ,
 \label{f_scal}
\end{align}
 with
 \begin{align}
 \mathcal{F}(p_F,q,{\bf p})=&\frac{m^2}{E({\bf p})E({\bf p+q})}\nonumber\\
 &\times \theta(p_F-|{\bf p}|)\theta(|{\bf p+q}|-p_F)\nonumber\\
 &\times \delta(\omega+E({\bf p})-E({\bf p+q}))\,
 \label{eq:scale_integrand}
 \end{align}
 resulting in~\cite{Alberico:1988bv,Donnelly:1991qy}
 \begin{align}
  \mathcal{F}(p_F,q,{\bf p}) =\frac{3\mathcal{N}m^2}{2p_F^3|{\bf q}|}\theta(E_F-\Gamma)(E_F-\Gamma)\ ,
\end{align}
 where we have introduced $E_F=\sqrt{p_F^2+m^2}$ and 
 \begin{align}
\Gamma&=\rm{Max}\{\Gamma_1,\Gamma_2,\Gamma_3\}\nonumber\\
&=\rm{Max}\Big\{m, E_F-\omega,\frac{-\omega+|{\bf q}|\sqrt{1+1/\tau}}{2}\Big\}\ .
 \end{align}
It is convenient to introduce the widespread set of dimensionless variables~\cite{Alberico:1988bv} 
\begin{align}
\lambda=\omega/2m\ ,\nonumber\\
\kappa=|{\bf q}|/2m\ ,\nonumber\\
\eta_F=p_F/m\ .
\end{align}
The minimum $\Gamma_3/m=1$ at
\begin{align}
\lambda=\lambda^0=\frac{1}{2}\Big[ \sqrt{(1+4 \kappa^2)}-1\Big]\ ,
\end{align} 
corresponds to the quasi elastic peak $\tau = \lambda$~\cite{Alberico:1988bv}. In the limit of large $|{\bf q}|$, the relation $\Gamma=\Gamma_3$ is satisfied for each value of $\omega$. Hence, a dimensionless scaling variable can be defined in terms of this quantity as~\cite{Alberico:1988bv}
\begin{align}
\psi=sign(\lambda-\lambda^0)\Big[\frac{1}{\xi_F}\Big(\frac{\Gamma_3}{m}-1\Big)\Big]^{1/2}\ ,
\end{align}
with $\xi_F= E_F/m -1$ and such that $\psi=0$ at the quasi elastic peak. Note that this definition of the scaling variable is equivalent to the more common expression 
\begin{align} 
\psi= \frac{1}{\sqrt{\xi_F}}\frac{\lambda-\tau}{\sqrt{(1+\lambda)\tau+\kappa\sqrt{\tau(1+\tau)}}} \,.
\end{align}

Collecting previous results one obtains
\begin{align}
F(p_F,q)=\frac{3\mathcal{N}\xi_F}{4 \eta_F^3 m \kappa}\big(1-\psi^2)\theta(1-\psi^2)\ .
\end{align}
Substituting Eq. \eqref{had:tens:FG} and \eqref{w12} into Eqs. \eqref{RL}, \eqref{RT} leads to the following expressions   for the L and T response functions
\begin{align}
R_L=&\frac{3 \mathcal{N} }{4\pi p_F^3}\int d^3p\ \mathcal{F}(p_F,q,{\bf p})\Big\{ -W_1(\tau)\Big(1-\frac{{\omega}^2}{q^2}\Big)\nonumber\\
&+\frac{W_2}{m^2}\Big[ E_p - \frac{p\cdot q}{q^2}\omega\Big]^2\Big\}\ ,\nonumber\\
R_T=&\frac{3 \mathcal{N} }{4\pi p_F^3}\int d^3p\ \mathcal{F}(p_F,q,{\bf p})\Big\{2 W_1(\tau)+\frac{W_2(\tau)}{m^2}{\bf p}_T^2\Big\}\ .
\end{align}
After performing the integrations, the responses can be cast in the form
\begin{align}
R_L=&  \frac{3\mathcal{N}\xi_F}{4 \eta_F^3 m \kappa}\big(1-\psi^2)\theta(1-\psi^2)\nonumber\\
&\times \Big\{\frac{\kappa^2}{\tau}[G_E^2(\tau)+W_2(\tau)\Delta ]\Big\}\ ,\nonumber\\
R_T=&  \frac{3\mathcal{N}\xi_F}{4 \eta_F^3 m \kappa}\big(1-\psi^2)\theta(1-\psi^2)\nonumber\\
&\times \Big\{ 2\tau G^2_M(\tau)+ W_2(\tau)\Delta\Big\}\ ,
\end{align}
where
\begin{align}
\Delta= \xi_F(1-\psi^2)\Big[\frac{\sqrt{\tau(1+\tau)}}{\kappa}+\xi_F(1-\psi^2)\frac{\tau}{3\kappa^2}\Big]\ .
\end{align}

The next step consists in the definition of the longitudinal and transverse scaling functions~\cite{Maieron:2001it} 
\begin{align}
f_L(\psi)= p_F\times \frac{R_L}{G_L}\ ,\nonumber\\
f_T(\psi)= p_F\times \frac{R_T}{G_T}\ ,
\end{align}
where 
\begin{align}
G_L=\frac{\mathcal{N}}{2\kappa} \Big\{\frac{\kappa^2}{\tau}[G_E^2(\tau)+W_2(\tau)\Delta ]\Big\}\nonumber\\
G_T=\frac{\mathcal{N}}{2\kappa}\Big\{ 2\tau G^2_M(\tau)+ W_2(\tau)\Delta\Big\}\ .
\label{FG:pre:fact}
\end{align}
Within the GRFG the same scaling function for the the longitudinal and transverse channel arises. This is
a symmetric function centered in $\psi=0$ 
\begin{align}
f(\psi)=f_L(\psi)= f_T(\psi)=\frac{3\xi_F}{2 \eta_F^2}\big(1-\psi^2)\theta(1-\psi^2)\ .
\end{align} 

In the non relativistic limit the L and T responses can be expressed as
\begin{align}
R_L=& \frac{3\mathcal{N}}{4\pi p_F^3}\int d^3p \frac{1}{2}\sum_{s,s^\prime}\Big\{ \chi^\dagger_s \rho^\dagger({\bf q})\chi_{s^\prime}\chi^\dagger_{s^\prime}\rho({\bf q})\chi_s\Big\}\nonumber\\
 &\times \theta(p_F-|{\bf p}|)\theta(|{\bf p+q}|-p_F)\nonumber\\
 &\times \delta\Big(\omega +\frac{{\bf p}^2}{2m}-\frac{|{\bf p+q}|^2}{2m}\Big)\ ,\nonumber\\
 R_T=& \frac{3\mathcal{N}}{4\pi p_F^3}\int d^3p \frac{1}{2}\sum_{s,s^\prime}\Big\{ \chi^\dagger_s {\bf j}_T^\dagger({\bf q})\chi_{s^\prime}\chi^\dagger_{s^\prime}{\bf j}_T({\bf q})\chi_s\Big\}\nonumber\\
 &\times \theta(p_F-|{\bf p}|)\theta(|{\bf p+q}|-p_F)\nonumber\\
 &\times \delta\Big(\omega +\frac{{\bf p}^2}{2m}-\frac{|{\bf p+q}|^2}{2m}\Big)\ ,
 \end{align}
where $s$ and $s^\prime$ are the spin quantum numbers of the nucleon in the initial and final state, respectively.

In the following, non relativistic scaling variable and functions are introduced with the same non relativistic reduction of the current operator and relativistic corrections as in the GFMC calculations~\cite{Carlson:2001mp}. Neglecting the very small spin-orbit relativistic correction in the definition of charge operator, the charge and current operators read
\begin{align}
\rho({\bf q})=&\frac{G_E(\tau)}{\sqrt{1+\tau}}\ ,\nonumber\\
{\bf j}_T({\bf q})=& \Big[ \frac{G_E(\tau)}{m}{\bf p}_T-i \frac{G_M(\tau)}{2m}{\bf q}\times {\bm \sigma}\Big]\ .
\end{align}
As opposed to the semi relativistic model of Ref.~\cite{Amaro:2006if}, in the GFMC relativistic corrections enter only in the current definition, while the kinematics is fully non relativistic.\\
In the non relativistic limit, Eq.~\eqref{f_scal} reduces to
\begin{align}
F^{nr}(p_F,q)= & \frac{3 \mathcal{N} }{4\pi p_F^3}\int d^3p\ \mathcal{F}^{nr}(p_F,q,{\bf p})\nonumber\\
= &\frac{3\mathcal{N}m^2}{2p_F^3|{\bf q}|}\theta(E^{nr}_F-\Gamma)(E^{nr}_F-\Gamma^{nr})\ ,
 \label{f_scal_nr}
 \end{align}
 with
 \begin{align}
 \mathcal{F}^{nr}(p_F,q,{\bf p})&= \theta(p_F-|{\bf p}|)\theta(|{\bf p+q}|-p_F)\nonumber\\
 &\times \delta\Big(\omega +\frac{{\bf p}^2}{2m}-\frac{|{\bf p+q}|^2}{2m}\Big)\ ,
 \end{align}
and
\begin{align}
\Gamma^{nr}&={\rm Max}\{\Gamma^{nr}_1,\Gamma^{nr}_2\}\nonumber\\
&={\rm Max}\Big\{ E^{nr}_F-\omega,m+\frac{1}{2m}\Big( \frac{\omega m}{|{\bf q}|}-\frac{|{\bf q}|}{2}\Big)^2\Big\}\ .
\label{eps_nr}
\end{align}
The non relativistic Fermi energy  reads $E^{nr}_F=m+ {p_F^2}/{2m}$. We can then introduce a non relativistic scaling variable given by
\begin{align}
\psi^{nr}= &\Big[\frac{1}{\xi^{nr}_F}\Big(\frac{\Gamma^{nr}}{m}-1\Big)\Big]^{1/2}= \frac{1}{\sqrt{2\xi^{nr}_F}}\Big(\frac{\lambda}{\kappa}-{\kappa}\Big)\ .
\end{align}
In the limit of large $|{\bf q}|$, Eq.~\eqref{f_scal_nr} can be written in terms of $\psi^{nr}$ as
\begin{align}
F^{nr}(p_F,q)=\frac{3\mathcal{N}\xi^{nr}_F}{4 \eta_F^3 m \kappa}\big(1-{\psi^{nr}}^2)\theta(1-{\psi^{nr}}^2)\ .
\end{align}

In analogy with the relativistic case, the longitudinal and transverse responses are expressed as
\begin{align}
R^{nr}_L&=\frac{3 \mathcal{N} }{4\pi p_F^3}\int d^3p\ \mathcal{F}^{nr}(p_F,q,{\bf p})\Big\{\frac{G^2_E(\tau)}{1+\tau}\Big\}\ \nonumber\\
&=\frac{3\mathcal{N}\xi_F}{4 \eta_F^3 m \kappa}\big(1-{\psi^{nr}}^2)\theta(1-{\psi^{nr}}^2) \Big\{\frac{G^2_E(\tau)}{1+\tau}\Big\}\ ,\\
R^{nr}_T&=\frac{3 \mathcal{N} }{4\pi p_F^3}\int d^3p\ \mathcal{F}^{nr}(p_F,q,{\bf p})\Big\{ \frac{G^2_E(\tau)}{m^2}p_T^2\nonumber\\
&+ \frac{G^2_M(\tau)}{2m^2}|{\bf q}|^2\Big\}\nonumber\\
&= \frac{3\mathcal{N}\xi_F}{4 \eta_F^3 m \kappa}\big(1-{\psi^{nr}}^2)\theta(1-{\psi^{nr}}^2)\nonumber\\
&\times\Big\{ G_E^2(\tau)\Delta^{nr}+ 2 G_M^2(\tau)\kappa^2\Big\}\ ,
\end{align}
where
\begin{align}
\Delta^{nr}= \xi^{nr}_F (1-{\psi^{nr}}^2)\ .
\end{align}
We then define the non relativistic longitudinal and transverse scaling functions as 
\begin{align}
f^{nr}_L(\psi^{nr})&= p_F\times \frac{R^{nr}_L}{G^{nr}_L}\ ,\nonumber\\
f^{nr}_T(\psi^{nr})&= p_F\times \frac{R^{nr}_T}{G^{nr}_T}\ ,
\end{align}
where
\begin{align}
G^{nr}_L&= \frac{\mathcal{N}}{2\kappa}\Big\{\frac{G^2_E(\tau)}{1+\tau}\Big\}\ ,\nonumber\\
G^{nr}_T&= \frac{\mathcal{N}}{2\kappa}\Big\{ G_E^2(\tau)\Delta^{nr}+ 2 G_M^2(\tau)\kappa^2\Big\} \ .
\label{g_nr}
\end{align}

In order to compare our results with the data, we introduce the experimental
scaling functions obtained from the extracted longitudinal and transverse responses for $^4$He and $^{12}$C
\begin{align}
f^{exp}_L&= p_F\times\frac{R^{exp}_L}{G_L}\ ,\nonumber\\
f^{exp}_T&= p_F\times \frac{R^{exp}_T}{G_T}\ .
\end{align}
It is long known that $f^{exp}_L$ clearly shows a scaling behavior in the limit
of large momentum transfer. On the other hand, sizable scaling violations occur in the transverse channel, due to  significant contributions given by
two-body currents, resonance excitations and inelastic scattering. Hence, the comparison with
 the experimental data will be performed considering only the longitudinal contribution, $f^{exp}_L$. 

 \section{Results}
\label{results}

%%%%%%%%%%%%%%%%%%%%%%%%%%%%%%%%%%%%%%%%%%%%%%%%%%%%%%%%%%
\begin{figure}[!t]
\centering
\includegraphics[scale=0.675]{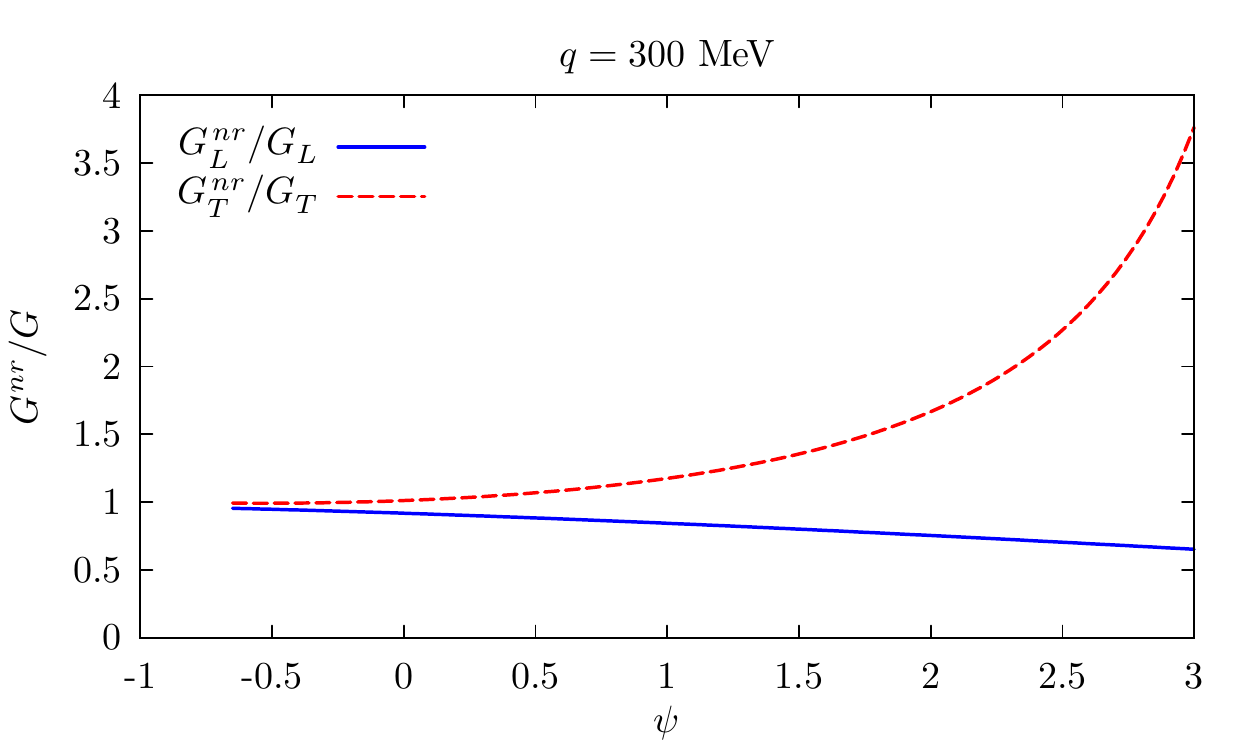}
\includegraphics[scale=0.675]{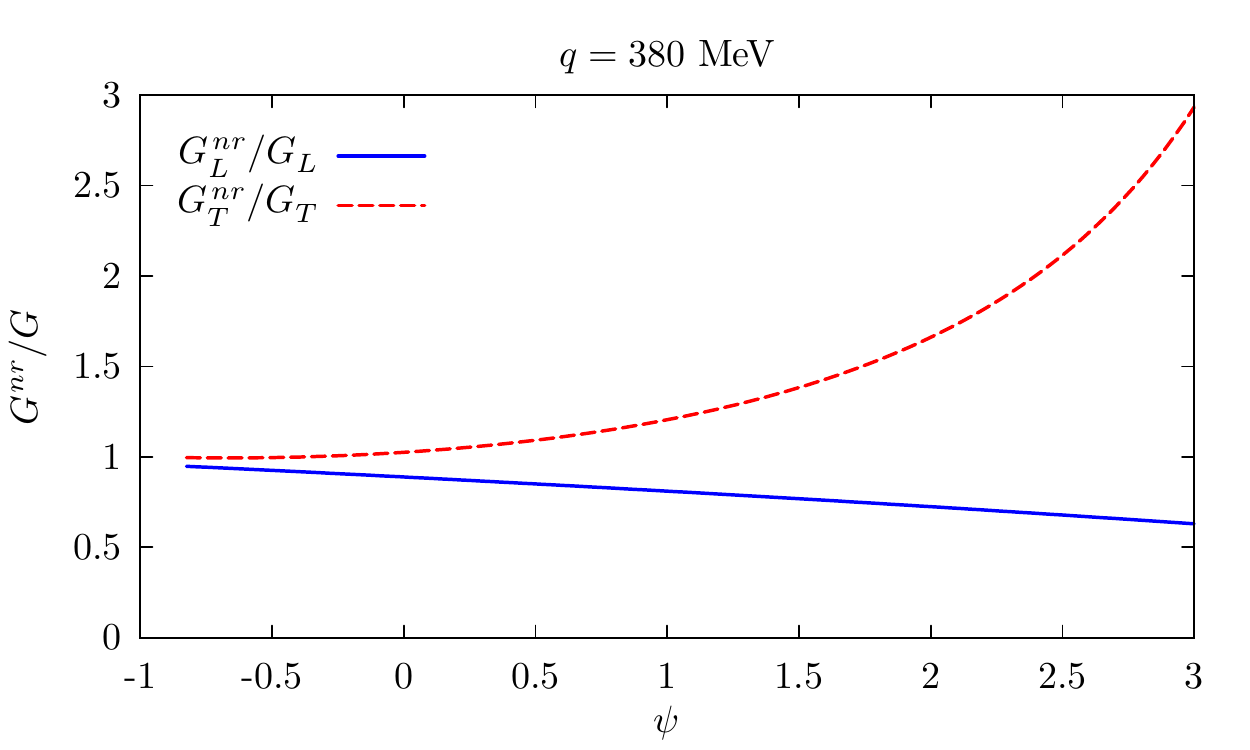}
\includegraphics[scale=0.675]{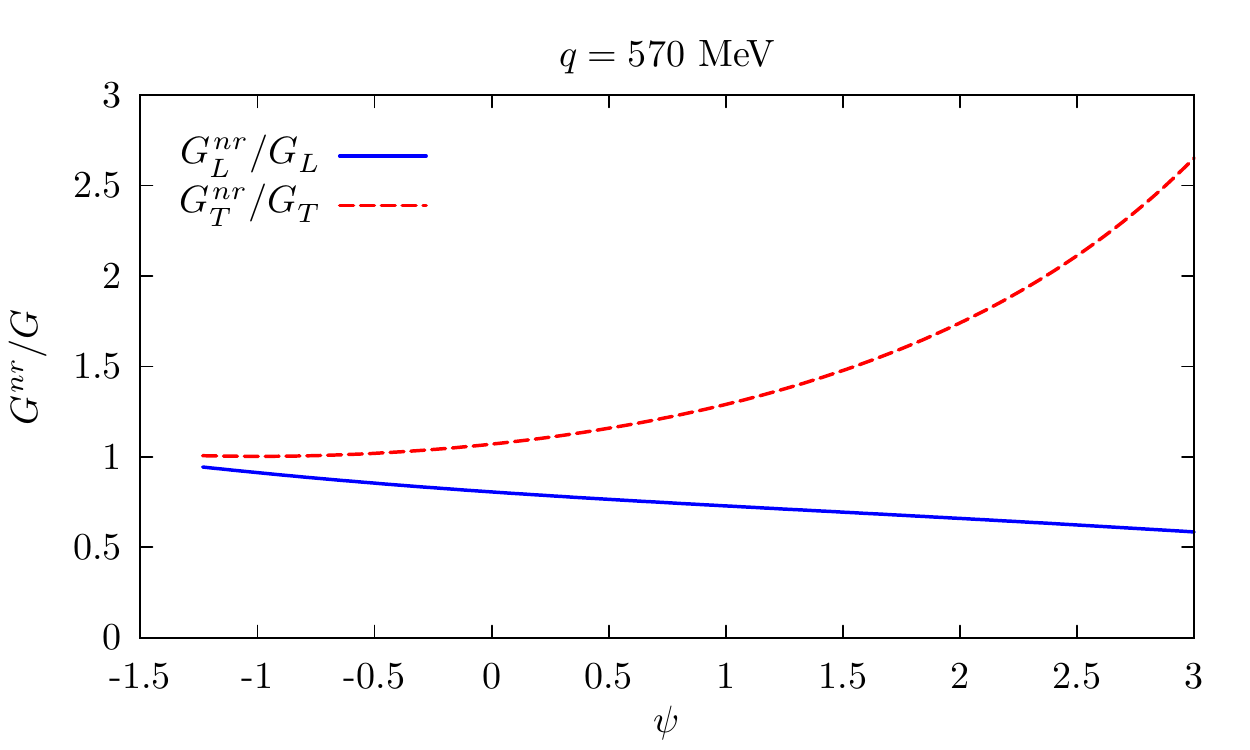}
\caption{(color online) Ratio of the non relativistic and relativistic expressions of the prefactors entering the definition of the scaling function plotted as a function of $\psi$, for $|{\bf q}|$= 300, 380, 570 MeV. 
The blue solid and red dashed lines correspond to the longitudinal and transverse channels, respectively. }
\label{prefact}
\end{figure}
%%%%%%%%%%%%%%%%%%%%%%%%%%%%%%%%%%%%%%%%%%%%%%%%%%%%%%%%%%

Here we analyze the scaling features of the GFMC responses. In order to highlight the underlying nuclear dynamics we first divide them by the non relativistic prefactors $G^{nr}_{L,T}$.  These have been obtained expanding the relativistic-current matrix elements in powers of $1/m$ retaining terms up to $\mathcal{O}[1/m^2]$ \cite{Lovato:2016gkq}. Relativistic corrections appear as terms of $\mathcal{O}[1/m^2]$ in the longitudinal channel while they are $\mathcal{O}[1/m^3]$ in  the transverse one, and are therefore neglected in this case. This difference plays a relevant role in the interpretation of the results presented below.
For a  meaningful comparison with the scaling functions extracted from experimental data,
we also present the results obtained using the relativistic prefactors $G_{L,T}$. 
Figure \ref{prefact} clearly shows the different behavior of $G^{nr}_{L,T}$ and $G_{L,T}$ for three values of the momentum transfer. Relativistic effects are particularly relevant in the transverse case; at $|{\bf q}|$= 570 MeV the ratio $G^{nr}_T/G_T$ significantly differs from 1 for $\psi\geq 0$.

%%%%%%%%%%%%%%%%%%%%%%%%%%%%%%%%%%%%%%%%%%%%%%%%%%%%%%%%%%
\begin{figure}[h]
\centering
\includegraphics[scale=0.675]{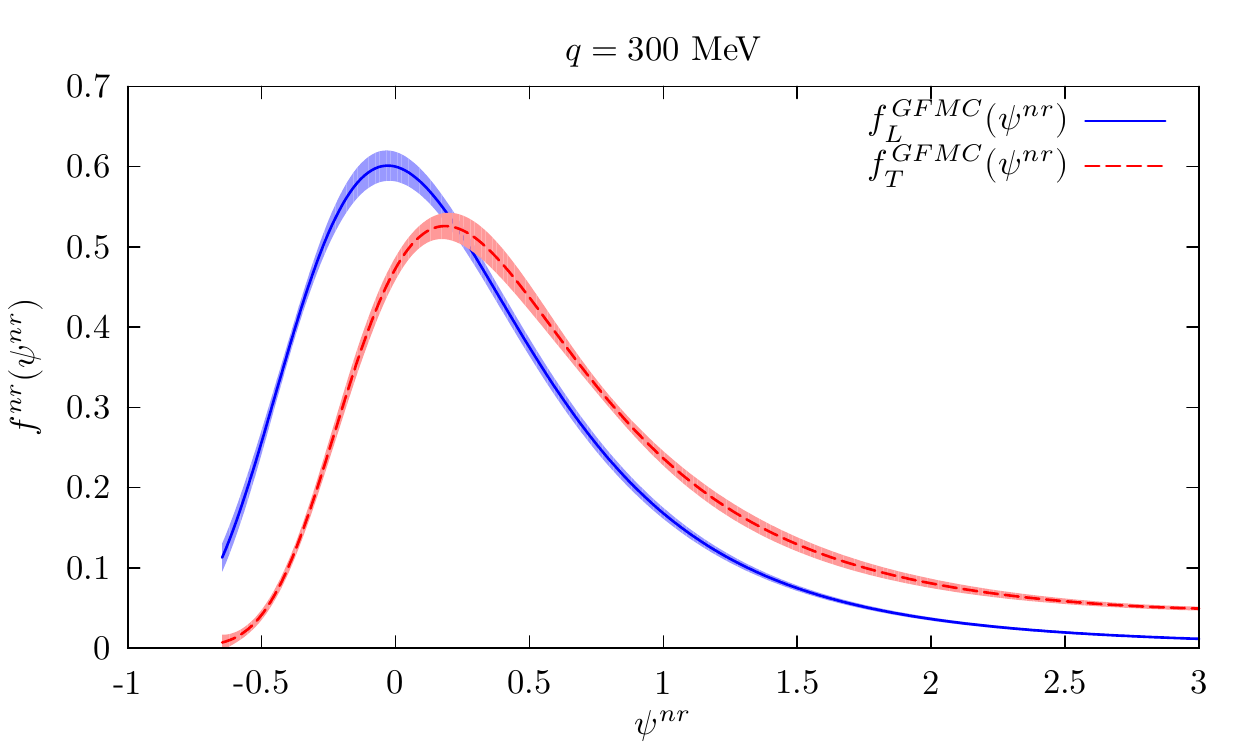}
\includegraphics[scale=0.675]{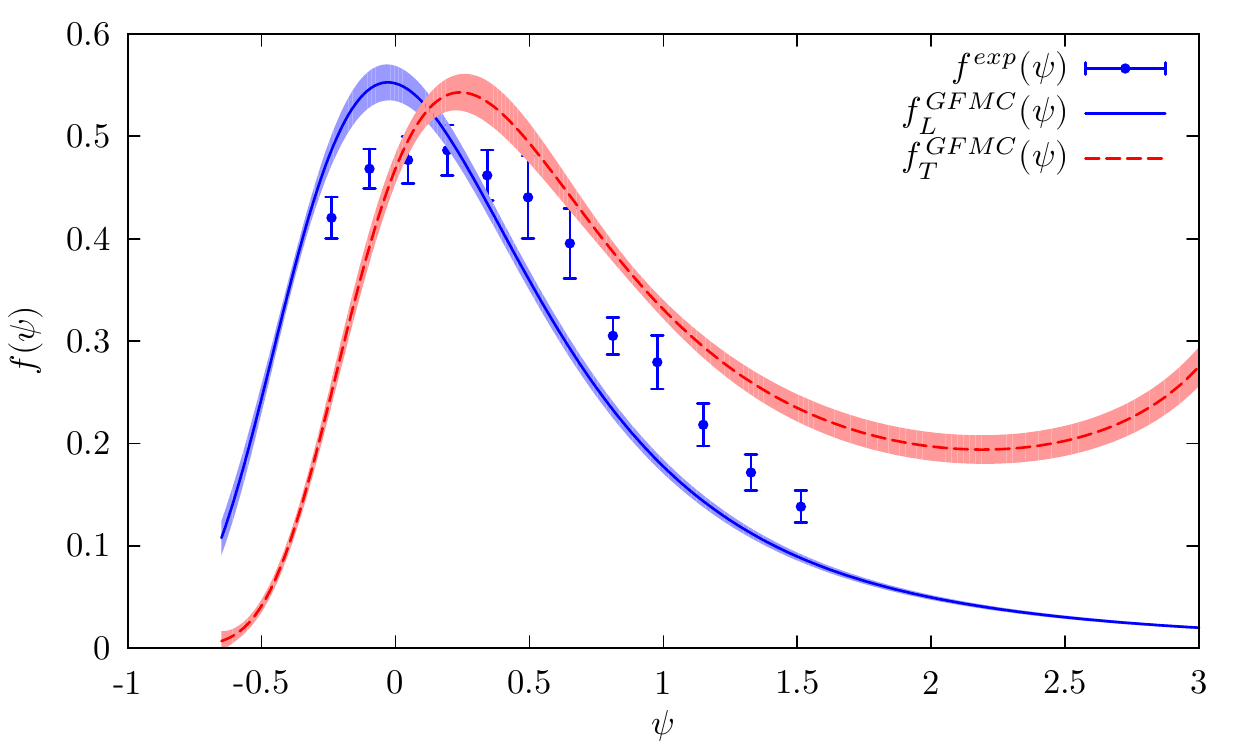}
\caption{(color online) Longitudinal (solid blue) and transverse (dashed red)  scaling functions obtained from the GFMC calculation of the longitudinal and transverse responses of $^{12}$C at  $|{\bf q}|= 300$ MeV. 
{\bf Upper panel}: the responses have been divided by the non relativistic prefactors and the resulting curves are plotted as a function of $\psi^{nr}$.
{\bf Lower panel}: the standard definition of the prefactors given in Eq.~\eqref{FG:pre:fact} has been used to get both the theoretical curves and the experimental points obtained from the data of Ref.~\cite{Barreau:1983ht} .
}
\label{300_12C}
\end{figure}
%%%%%%%%%%%%%%%%%%%%%%%%%%%%%%%%%%%%%%%%%%%%%%%%%%%%%%%%%%

%%%%%%%%%%%%%%%%%%%%%%%%%%%%%%%%%%%%%%%%%%%%%%%%%%%%%%%%%%
\begin{figure}[]
\centering
\includegraphics[scale=0.675]{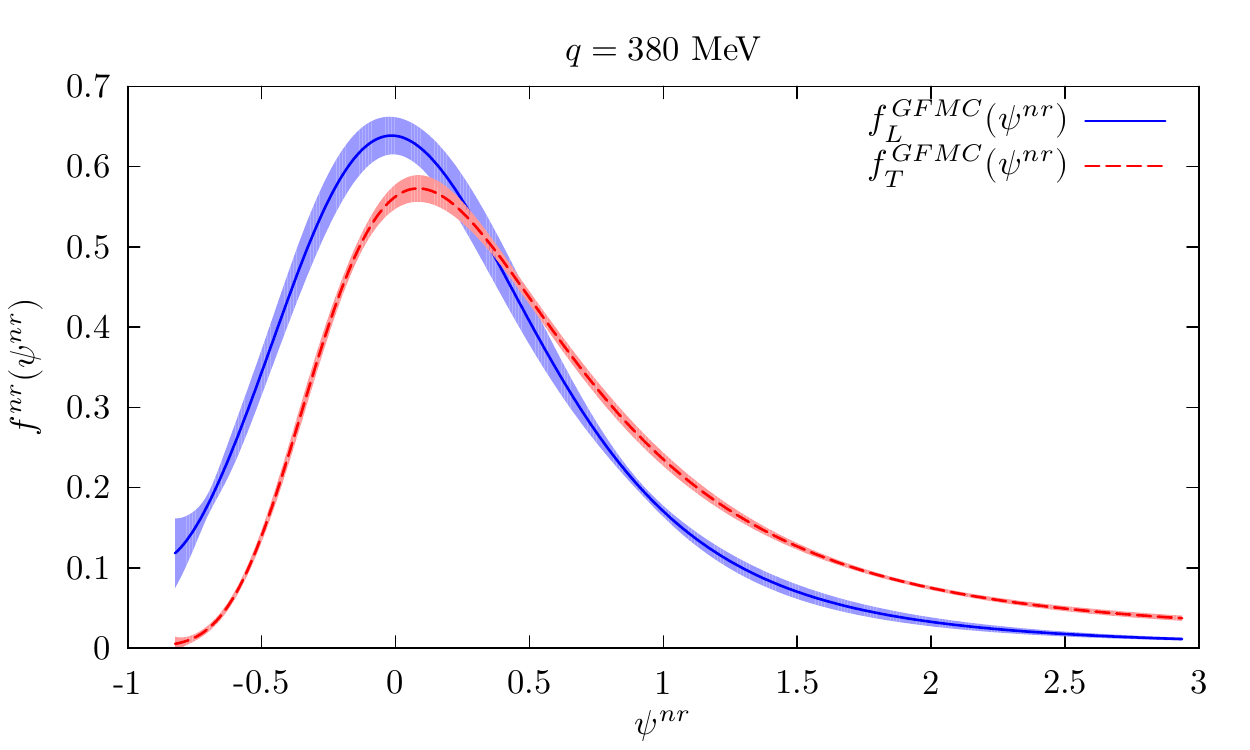}
\includegraphics[scale=0.675]{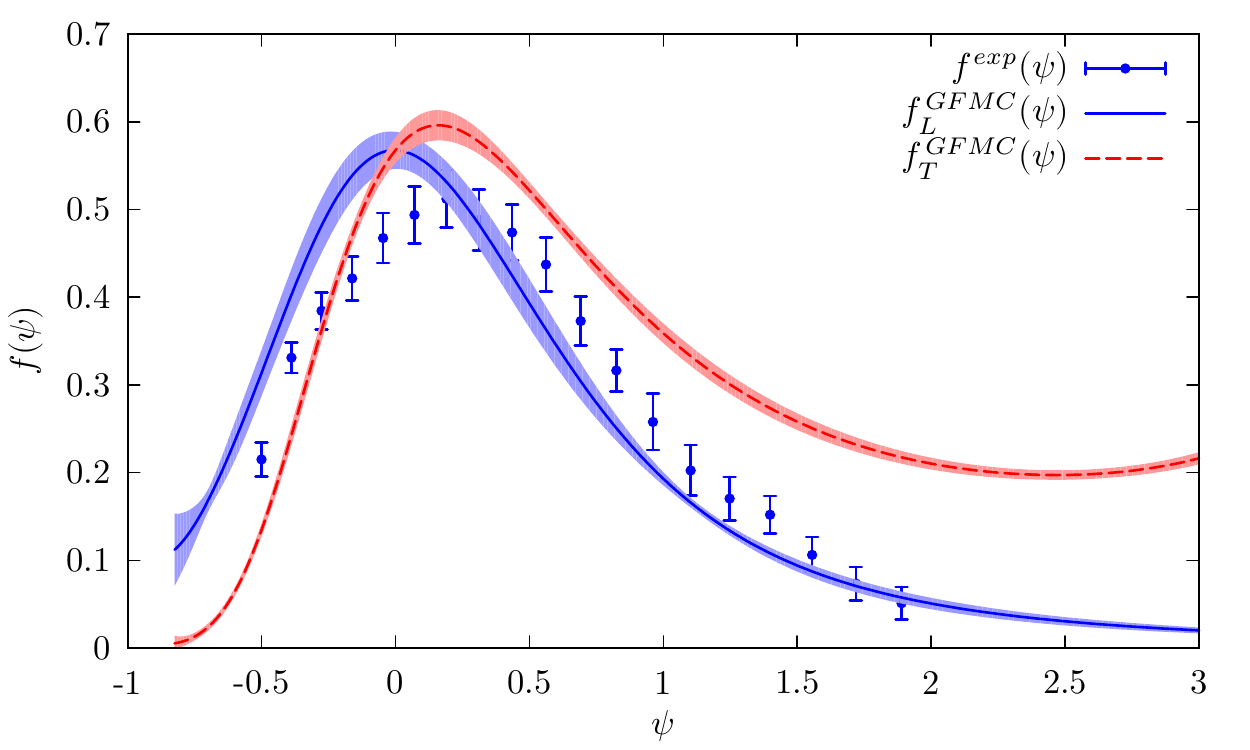}
\caption{ Same as in Fig. \ref{300_12C} but for  $|{\bf q}|= 380$ MeV.}
\label{380_12C}
\end{figure}
%%%%%%%%%%%%%%%%%%%%%%%%%%%%%%%%%%%%%%%%%%%%%%%%%%%%%%%%%%

%%%%%%%%%%%%%%%%%%%%%%%%%%%%%%%%%%%%%%%%%%%%%%%%%%%%%%%%%%
\begin{figure}[]
\includegraphics[scale=0.675]{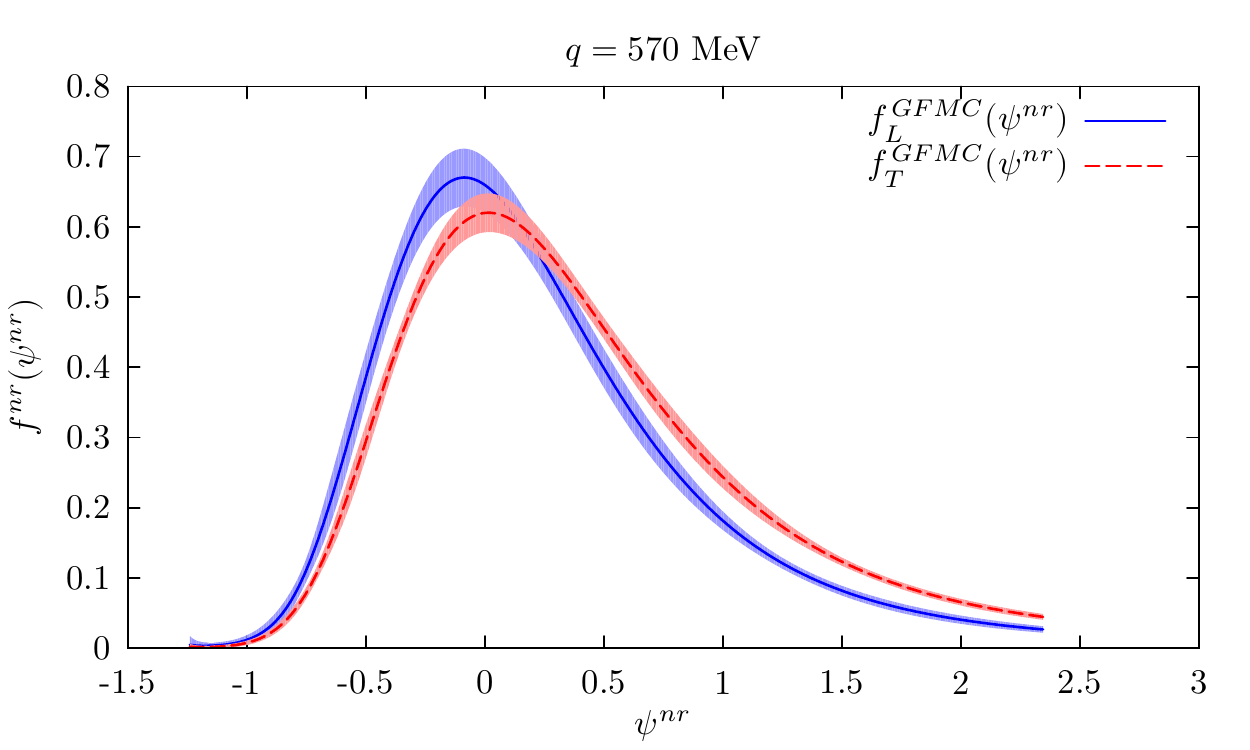}
\includegraphics[scale=0.675]{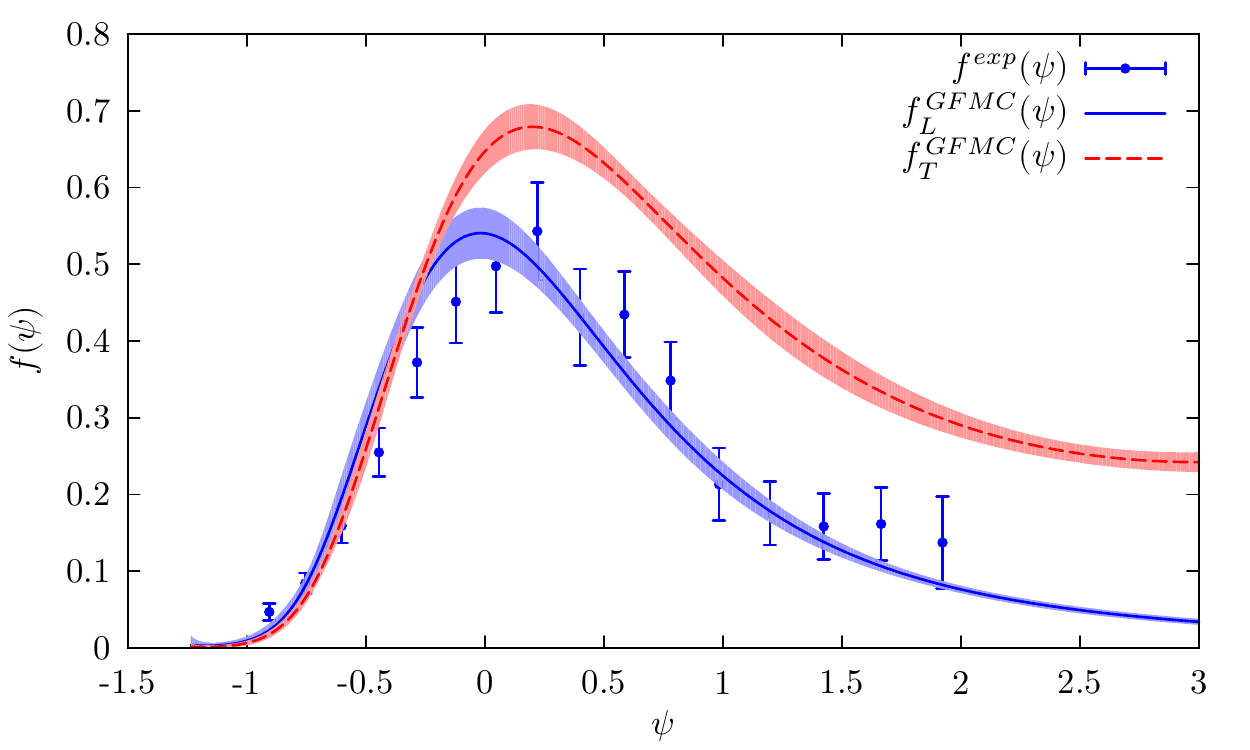}
\caption{ Same as in Fig. \ref{300_12C} but for  $|{\bf q}|= 570$ MeV.}
\label{570_12C}
\end{figure}
%%%%%%%%%%%%%%%%%%%%%%%%%%%%%%%%%%%%%%%%%%%%%%%%%%%%%%%%%%

In Figs.~\ref{300_12C}, \ref{380_12C} and \ref{570_12C} we show the longitudinal (blue solid lines) and 
transverse (red dashed lines) scaling functions extracted from the GFMC calculations of the $^{12}$C response functions. The results in the upper panels, obtained dividing the GFMC calculations by $G^{nr}_{L(T)}$, are plotted as a function of the non relativistic scaling variable $\psi^{nr}$. In the lower panels a comparison between the theoretical curves and the experimental points, in which the relativistic form of the prefactors has been adopted, is presented. 

 It is important to point out that the longitudinal response of $^{12}$C is known to be affected by the elastic and the low lying excited states\textemdash $J^{\pi}=2^+,\ 0_2^+,$ and $4^+$\textemdash contributions. In order to compare experiments \textemdash which refer only to the inclusive quasi-elastic response\textemdash with GFMC calculations,
these contributions have been explicitly subtracted by using the experimental values of excitations energies and form factors. 
Because of the fast drop of the form factors with increasing momentum transfer, in Ref. \cite{Lovato:2016gkq} it is argued that these corrections are expected to be significant in the longitudinal channel at $|{\bf q}|= 300$ MeV, but almost negligible at $|{\bf q}|= 570$ MeV. On the other hand, in the transverse channel such contributions are expected to be always negligible.

The scaling functions displayed in the upper panels exhibit a clearly asymmetric shape, with a
tail extending in the region $\psi^{nr} > 0$, as opposed to the GRFG model predictions. The difference in magnitude between the longitudinal and transverse GFMC scaling functions, which become less evident for larger values of $|{\bf q}|$, is likely to be ascribed to small residual effects of the low lying excited state contributions. For the aforementioned reason, in the lower panels the agreement between the longitudinal GFMC scaling function and  the experimental data improves with increasing momentum transfer. 

The different behavior of the transverse scaling functions displayed in the upper and lower panels deserves some comments. In the lower panels, the red curves present a large non vanishing tail for $\psi >1$, although those are expected to approach zero, as shown in the upper panels. This discrepancy can be best understood considering the  results of Fig.\ref{prefact}. The relativistic and non relativistic expressions of the transverse prefactors used to extract the scaling functions are sizably different in the kinematic setups considered. In particular, for $|{\bf q}|=570$ MeV, these are very similar for $-1.5\leq  \psi \leq 0$ where their ratio is almost 1, while in the region $\psi\geq 0$ their trend is significantly different and $G^{nr}_T/G_T$ increases for larger values of $\psi$.

%%%%%%%%%%%%%%%%%%%%%%%%%%%%%%%%%%%%%%%%%%%%%%%%%%%%%%%%%%
\begin{figure}[]
\includegraphics[scale=0.675]{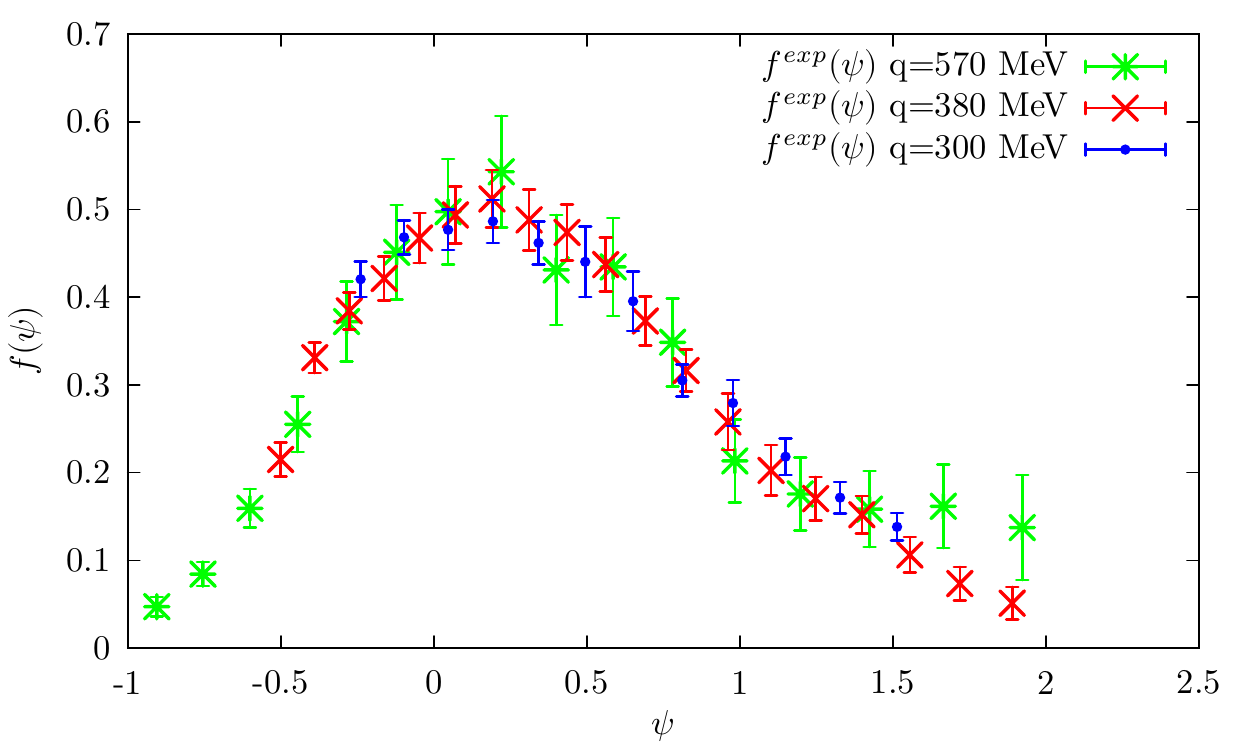}
\vspace*{-.1in}
\caption{(color online) Experimental scaling functions of $^{12}$C obtained from the longitudinal responses for $|{\bf q}|=300,\ 380,\ 570$ MeV \cite{Barreau:1983ht}.  }
\label{fl_exp_12C}
\end{figure}
%%%%%%%%%%%%%%%%%%%%%%%%%%%%%%%%%%%%%%%%%%%%%%%%%%%%%%%%%%

%%%%%%%%%%%%%%%%%%%%%%%%%%%%%%%%%%%%%%%%%%%%%%%%%%%%%%%%%%
\begin{figure}[]
\includegraphics[scale=0.675]{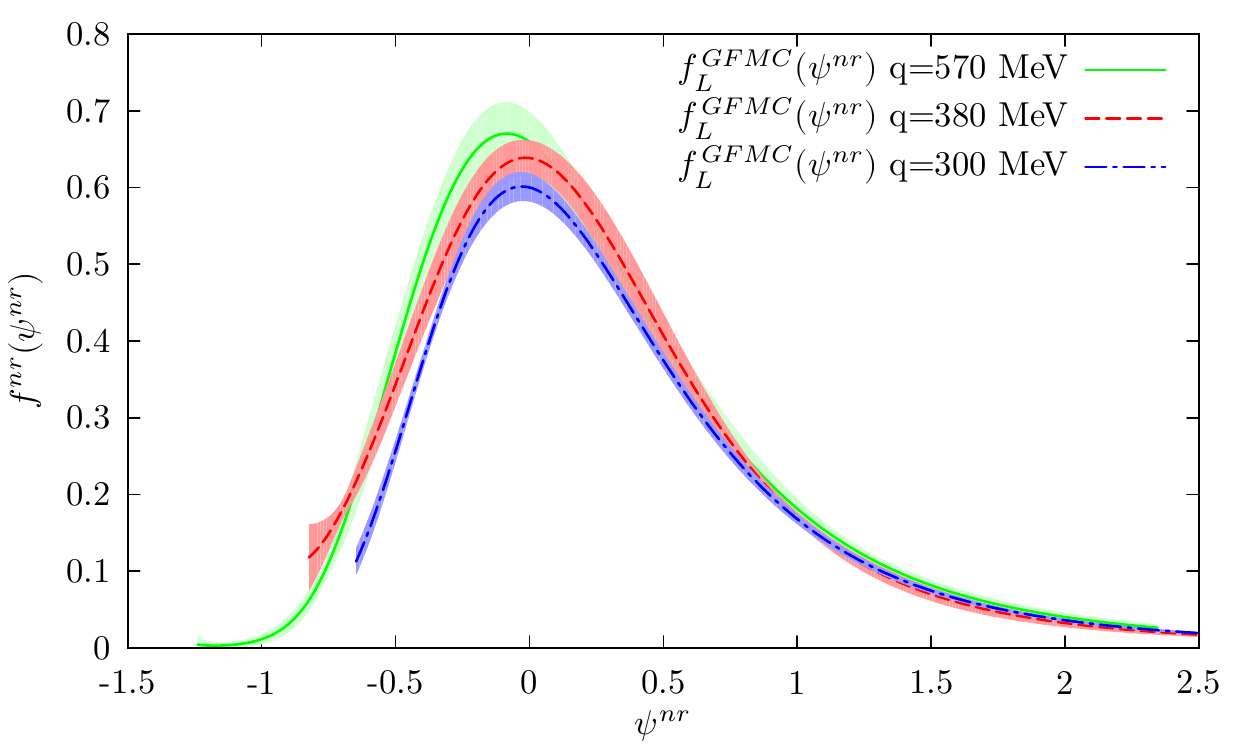}
\vspace*{-.1in}
\caption{(color online) Longitudinal scaling functions  of $^{12}$C obtained from GFMC calculations for $|{\bf q}|=300,\ 380,\ 570$ MeV as a function of $\psi^{nr}$.  }
\label{fl_all_12C}
\end{figure}
%%%%%%%%%%%%%%%%%%%%%%%%%%%%%%%%%%%%%%%%%%%%%%%%%%%%%%%%%%

Figure \ref{fl_exp_12C} shows the experimental scaling functions of $^{12}$C extracted from the experimental data of Ref. \cite{Barreau:1983ht} for $|{\bf q}|=300,\ 380,$ and 570 MeV. Although scaling is expected to occur in the limit of large momentum transfer, within the error bars of the different data points, the longitudinal response functions scale to a universal curve over the entire quasi-elastic peak, even in the region of moderate $|{\bf q}|$.

In Fig. \ref{fl_all_12C} the longitudinal GFMC scaling functions are shown as a function of $\psi^{nr}$ for $|{\bf q}|=300,\ 380,$ and 570 MeV. 
The theoretical results seem to indicate that first-kind scaling occurs. However, the interpretation of the differences between the three curves is obscured by the residual effect of the low-lying transitions discussed above. A more meaningful comparison can be carried out in the transverse channel, where the response functions are not affected by this effect.

%%%%%%%%%%%%%%%%%%%%%%%%%%%%%%%%%%%%%%%%%%%%%%%%%%%%%%%%%%
\begin{figure}[]
\includegraphics[scale=0.675]{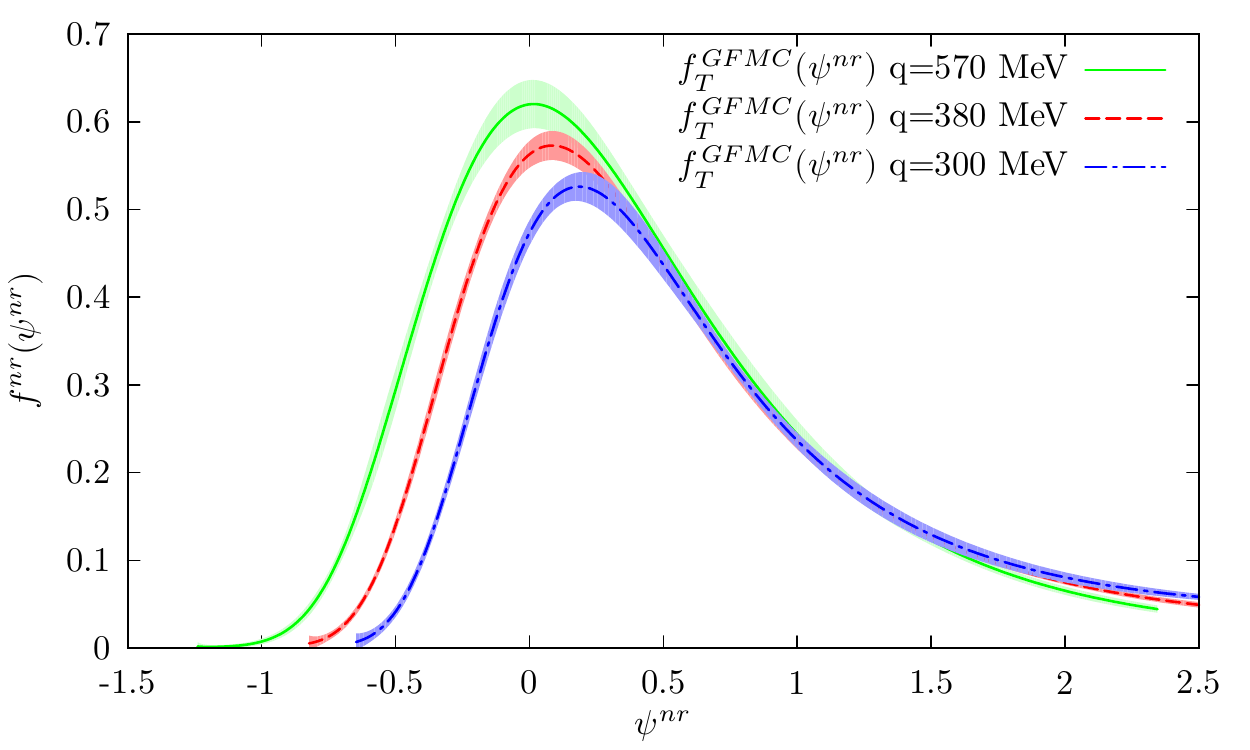}
\vspace*{-.1in}
\caption{(color online) Transverse scaling functions of $^{12}$C obtained from GFMC calculations for $|{\bf q}|=300,\ 380,\ 570$ MeV as a function of $\psi^{nr}$.  }
\label{ft_all_12C}
\end{figure}
%%%%%%%%%%%%%%%%%%%%%%%%%%%%%%%%%%%%%%%%%%%%%%%%%%%%%%%%%%

Figure \ref{ft_all_12C} shows the GFMC results for the transverse scaling functions. The difference between the three curves in the ${\psi}^{nr}<0$ region suggests that, for $|{\bf q}|=300,\ 380$ MeV, the requirement $\Gamma=\Gamma_2$ [see Eq.~\eqref{eps_nr}]\textemdash which is necessary to introduce the scaling variable\textemdash  is not satisfied for all the values of $\omega$. Indeed, the scaling violation in the low-energy transfer region is clearly visible. 

%%%%%%%%%%%%%%%%%%%%%%%%%%%%%%%%%%%%%%%%%%%%%%%%%%%%%%%%%%
\begin{figure}[]
\includegraphics[scale=0.675]{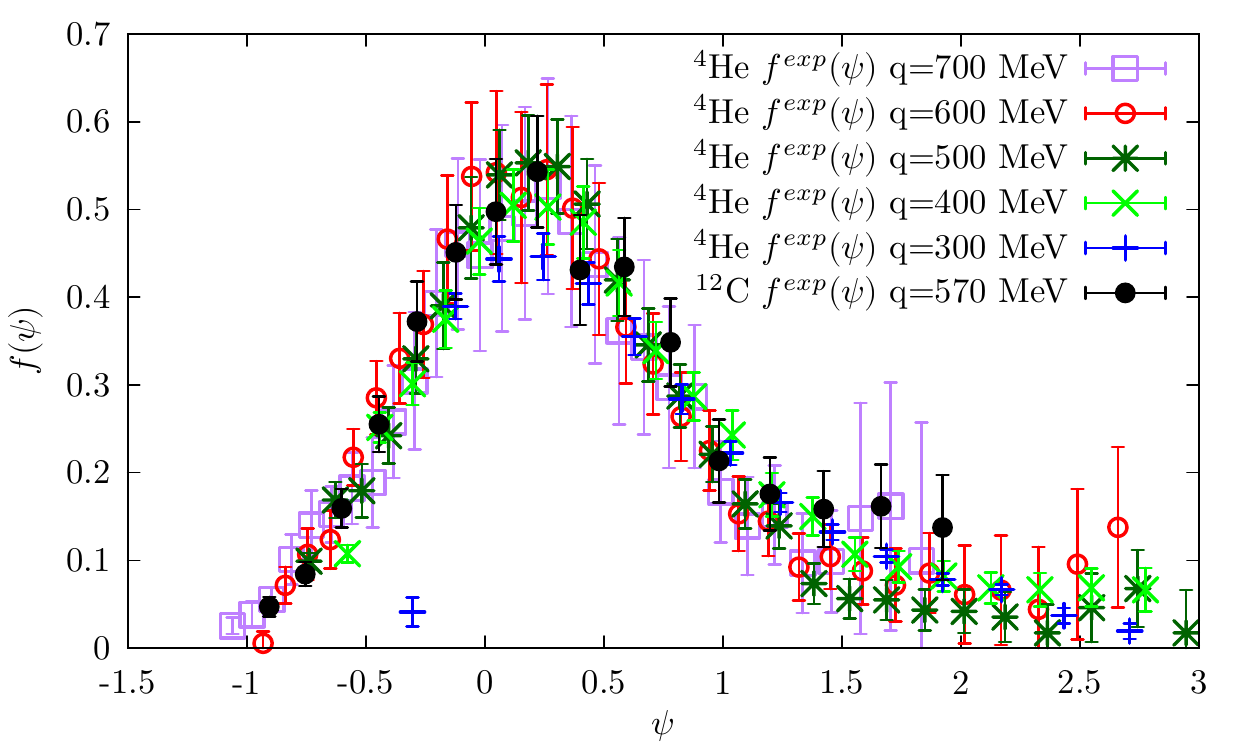}
\vspace*{-.1in}
\caption{(color online) Experimental scaling functions obtained from the longitudinal responses of $^4$He for $|{\bf q}|=$300, 400, 500, 600 and 700~MeV~\cite{Carlson:2001mp}. The value of the Fermi momentum of $^4$He has been set  to $180$ MeV. The black dots correspond to the scaling function obtained from the experimental longitudinal response of $^{12}$C at $|{\bf q}|=570$ MeV \cite{Barreau:1983ht}. }
\label{fl_exp_4He}
\end{figure}
%%%%%%%%%%%%%%%%%%%%%%%%%%%%%%%%%%%%%%%%%%%%%%%%%%%%%%%%%%

To better elucidate the scaling properties of the GFMC calculations, it is worth to analyze the  $^4$He nucleus, whose longitudinal response functions are not affected by low-lying transitions. In Fig.~\ref{fl_exp_4He}, the scaling functions obtained from the experimental data of the longitudinal responses of $^4$He at $|{\bf q}|= 300,\ 400,\ 500\ ,600,$ and $700$ MeV are shown. Choosing the Fermi momentum equal to $180$ MeV, we observe that the points corresponding to different values of the momentum transfer tend to lay on top of each other, and the agreement with the $^{12}$C 
data at $|{\bf q}|= 570$ MeV is also remarkable.

%%%%%%%%%%%%%%%%%%%%%%%%%%%%%%%%%%%%%%%%%%%%%%%%%%%%%%%%%%
\begin{figure}[]
\centering
\includegraphics[scale=0.675]{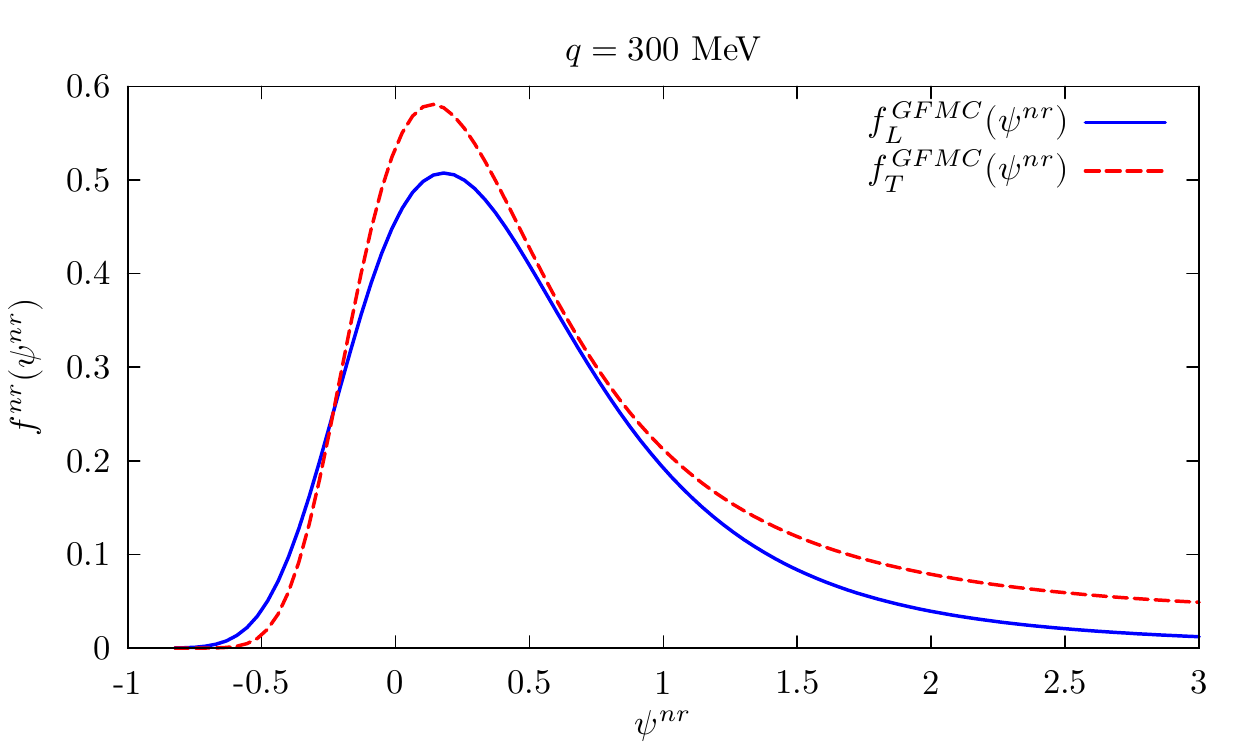}
\includegraphics[scale=0.675]{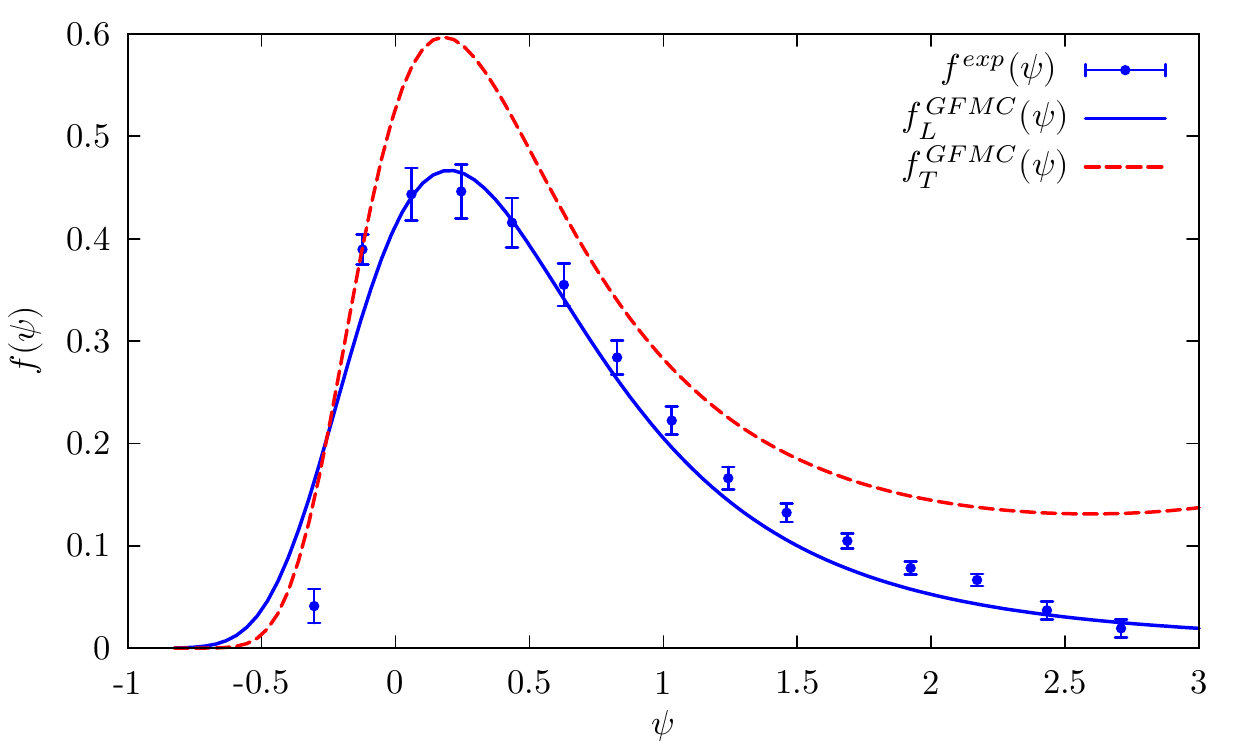}
\caption{(color online) Longitudinal (solid blue) and transverse (dashed red)  scaling functions obtained from the GFMC calculation of the longitudinal and transverse responses of $^{4}$He at  $|{\bf q}|= 300$ MeV. 
{\bf Upper panel}: the responses have been divided by the non relativistic prefactors and the resulting curves are plotted as a function of $\psi^{nr}$.
{\bf Lower panel}: the standard definition of the prefactors given in Eq.\eqref{FG:pre:fact} has been used to get both the theoretical curves and the experimental points obtained from the data of Ref. \cite{Carlson:2001mp} .}
\label{300_4He}
\end{figure}
%%%%%%%%%%%%%%%%%%%%%%%%%%%%%%%%%%%%%%%%%%%%%%%%%%%%%%%%%%

%%%%%%%%%%%%%%%%%%%%%%%%%%%%%%%%%%%%%%%%%%%%%%%%%%%%%%%%%%
\begin{figure}[]
\centering
\includegraphics[scale=0.675]{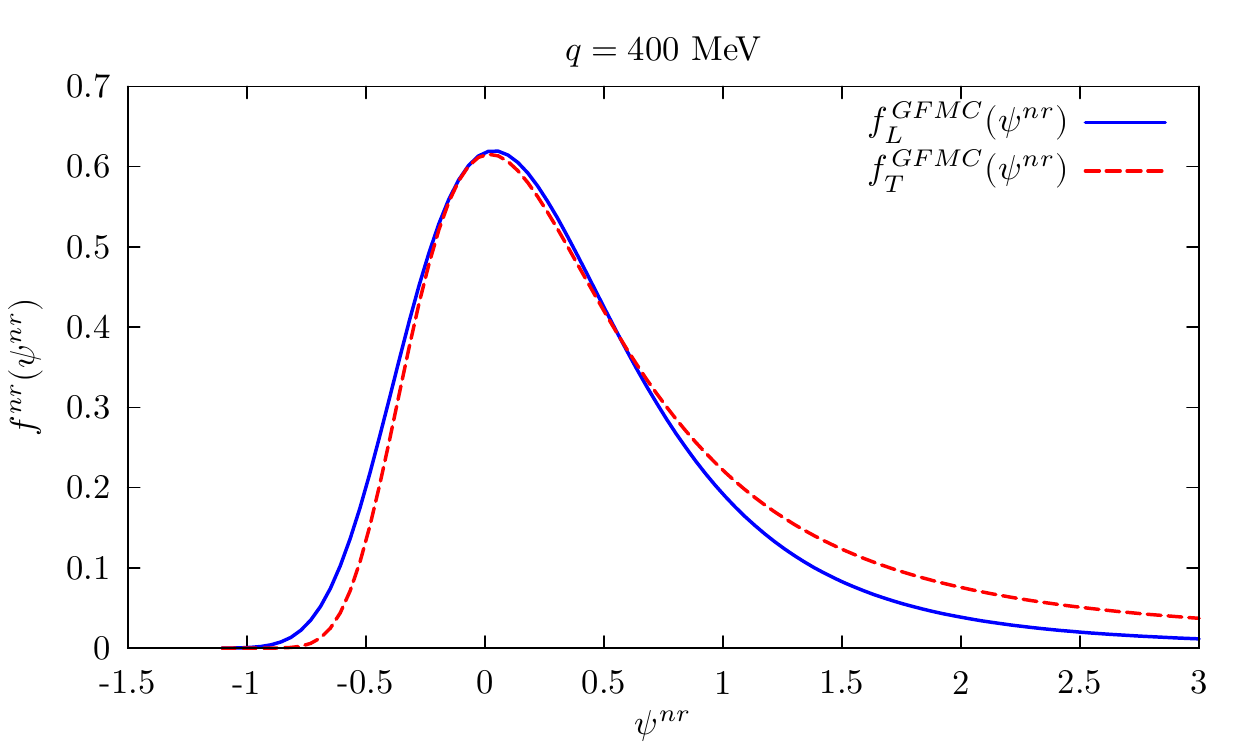}
\includegraphics[scale=0.675]{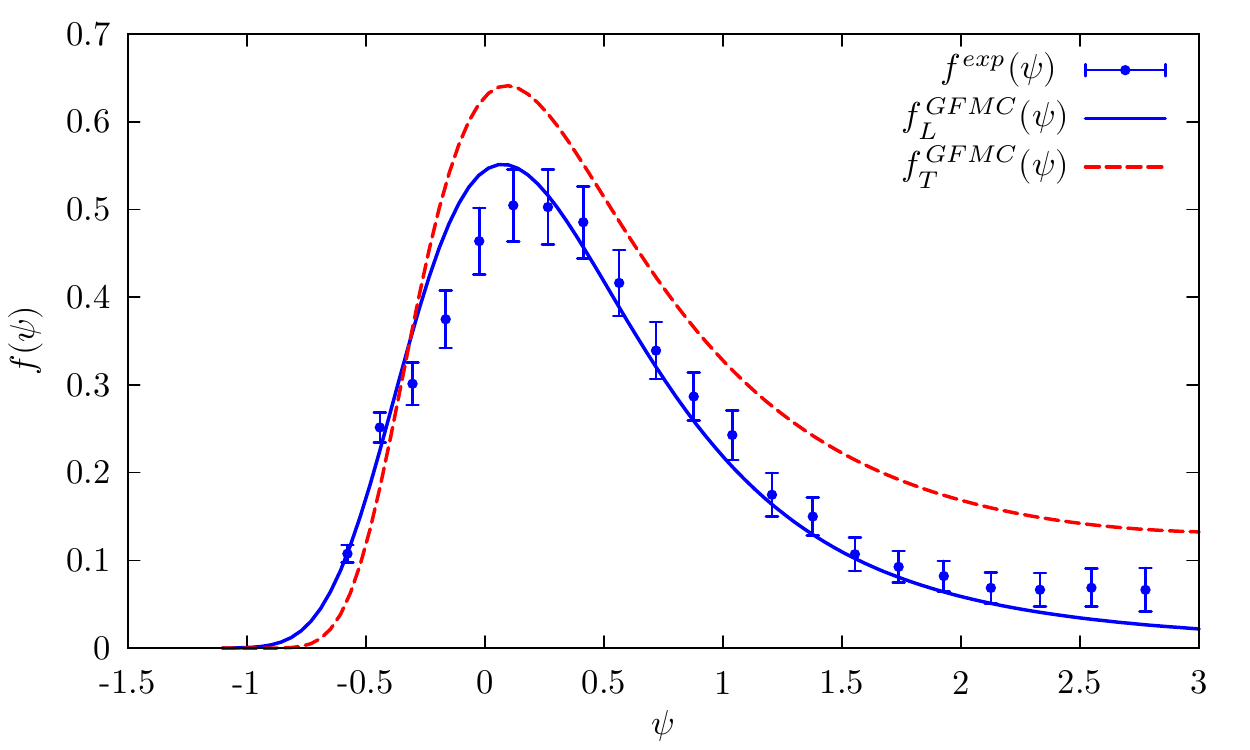}
\caption{ Same as in Fig. \ref{300_4He} but for  $|{\bf q}|= 400$ MeV.}
\label{400_4He}
\end{figure}
%%%%%%%%%%%%%%%%%%%%%%%%%%%%%%%%%%%%%%%%%%%%%%%%%%%%%%%%%%

%%%%%%%%%%%%%%%%%%%%%%%%%%%%%%%%%%%%%%%%%%%%%%%%%%%%%%%%%%
\begin{figure}[]
\includegraphics[scale=0.675]{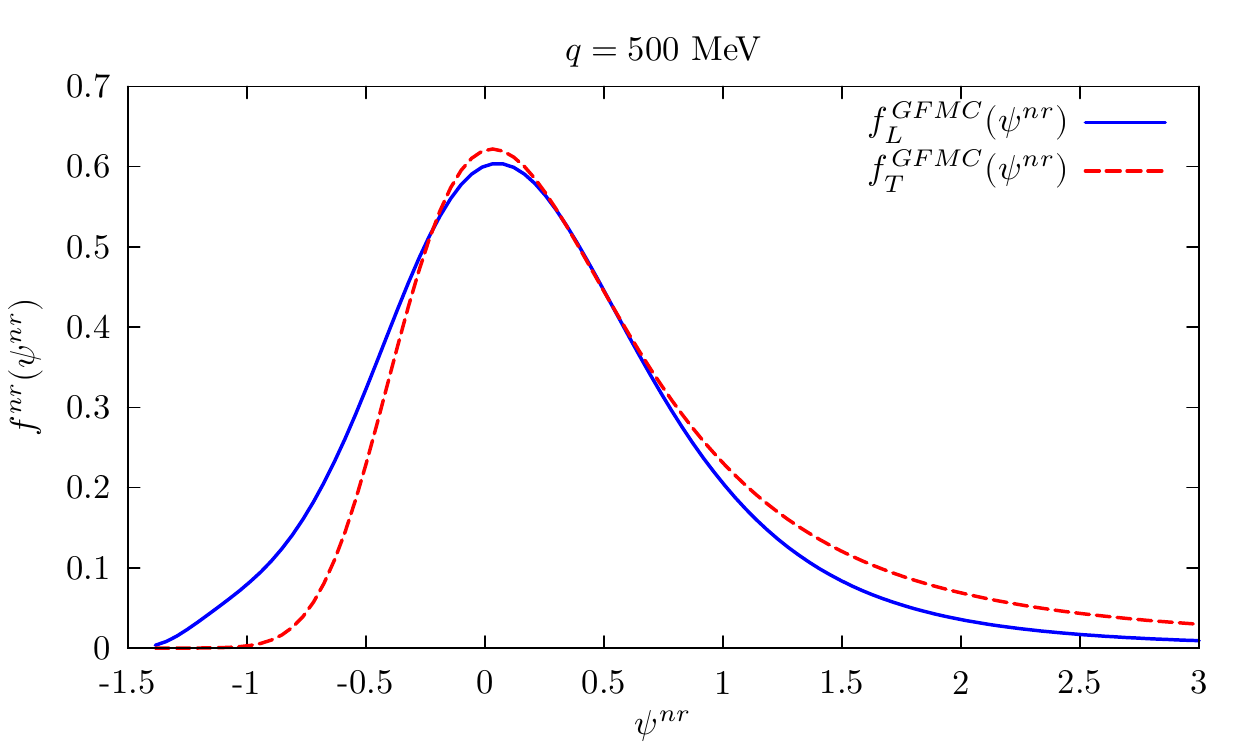}
\includegraphics[scale=0.675]{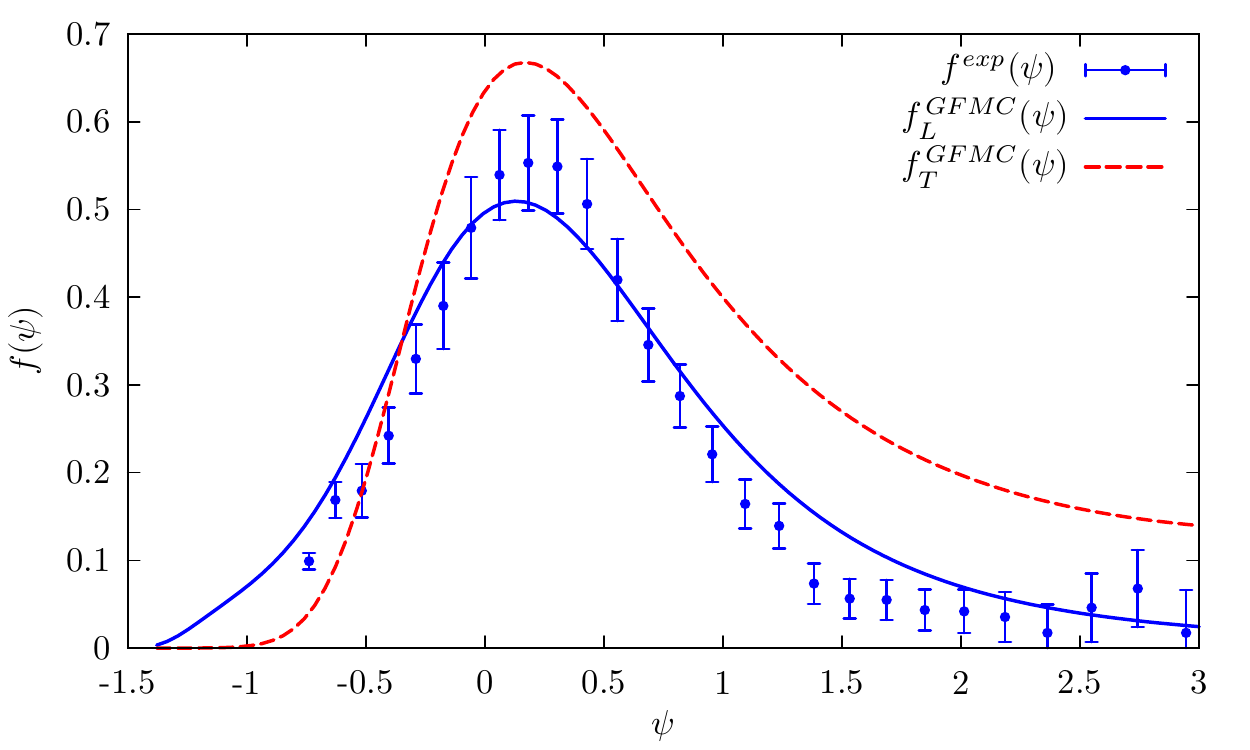}
\caption{ Same as in Fig. \ref{300_4He}  but for  $|{\bf q}|= 500$ MeV.}
\label{500_4He}
\end{figure}
%%%%%%%%%%%%%%%%%%%%%%%%%%%%%%%%%%%%%%%%%%%%%%%%%%%%%%%%%%

%%%%%%%%%%%%%%%%%%%%%%%%%%%%%%%%%%%%%%%%%%%%%%%%%%%%%%%%%%
\begin{figure}[]
\includegraphics[scale=0.675]{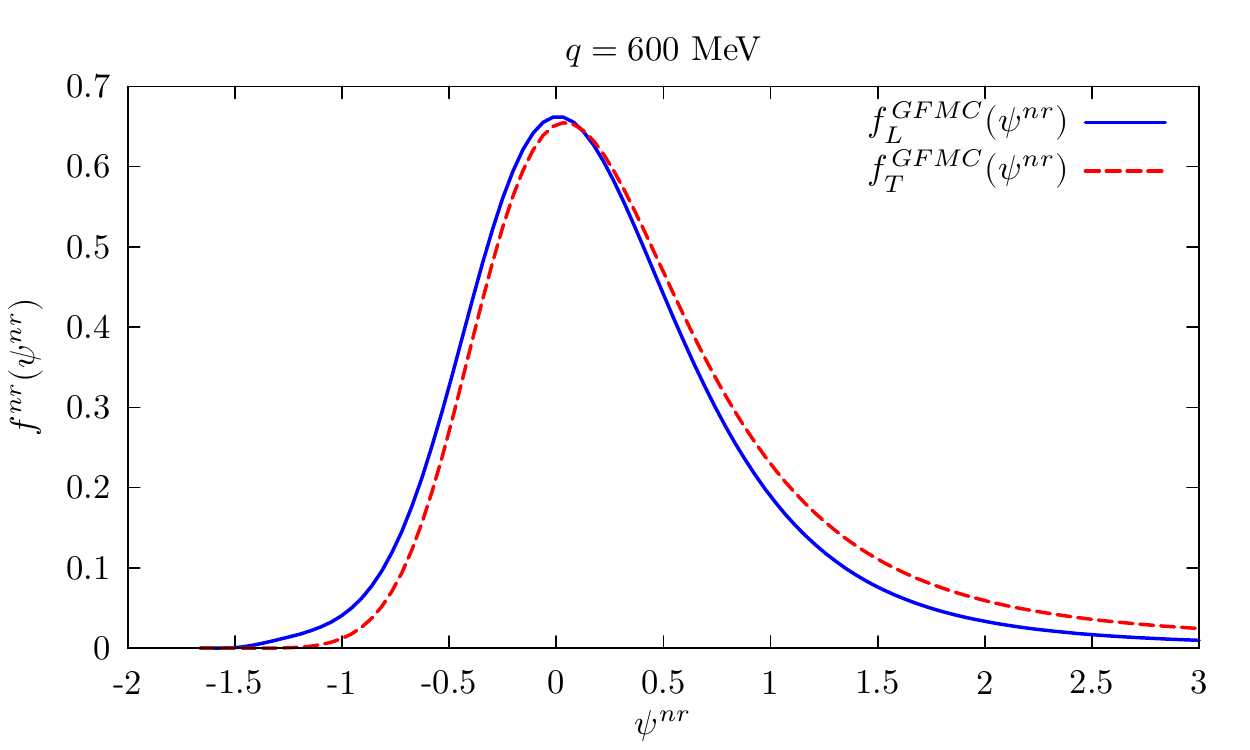}
\includegraphics[scale=0.675]{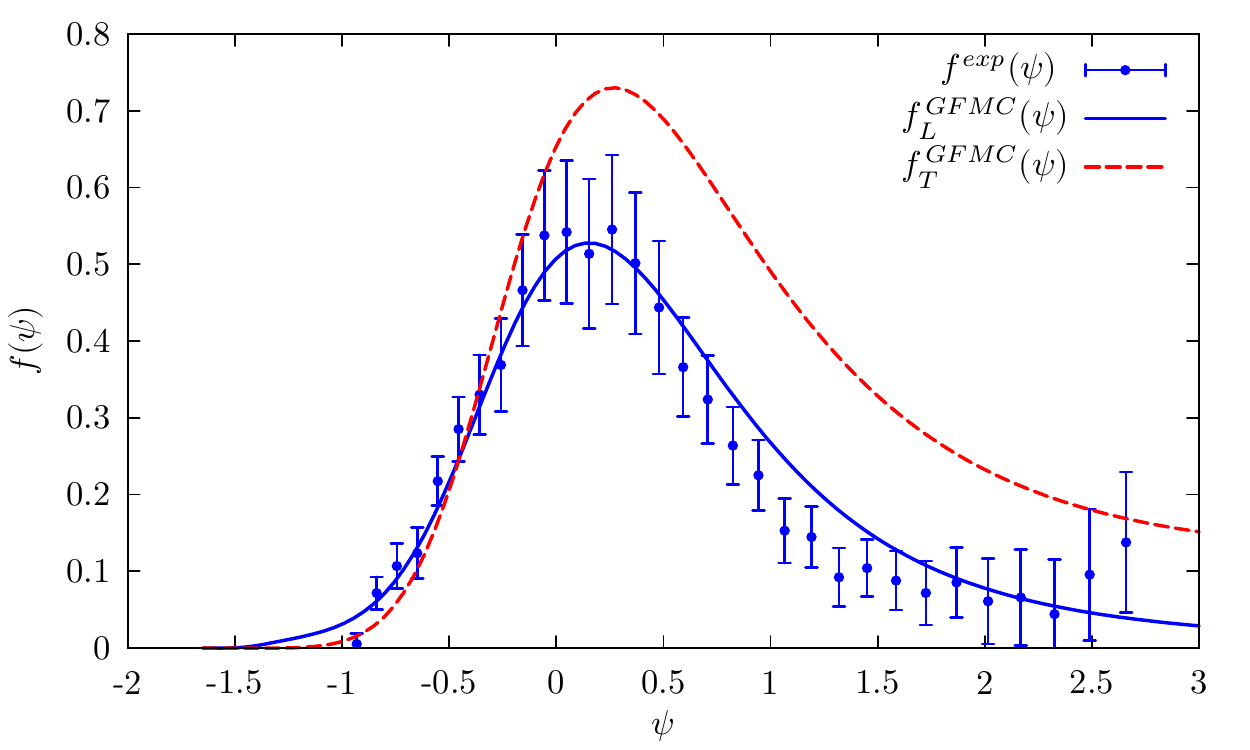}
\caption{ Same as in Fig. \ref{300_4He}  but for  $|{\bf q}|= 600$ MeV.}
\label{600_4He}
\end{figure}
%%%%%%%%%%%%%%%%%%%%%%%%%%%%%%%%%%%%%%%%%%%%%%%%%%%%%%%%%%
%%%%%%%%%%%%%%%%%%%%%%%%%%%%%%%%%%%%%%%%%%%%%%%%%%%%%%%%%%
\begin{figure}[]
\includegraphics[scale=0.675]{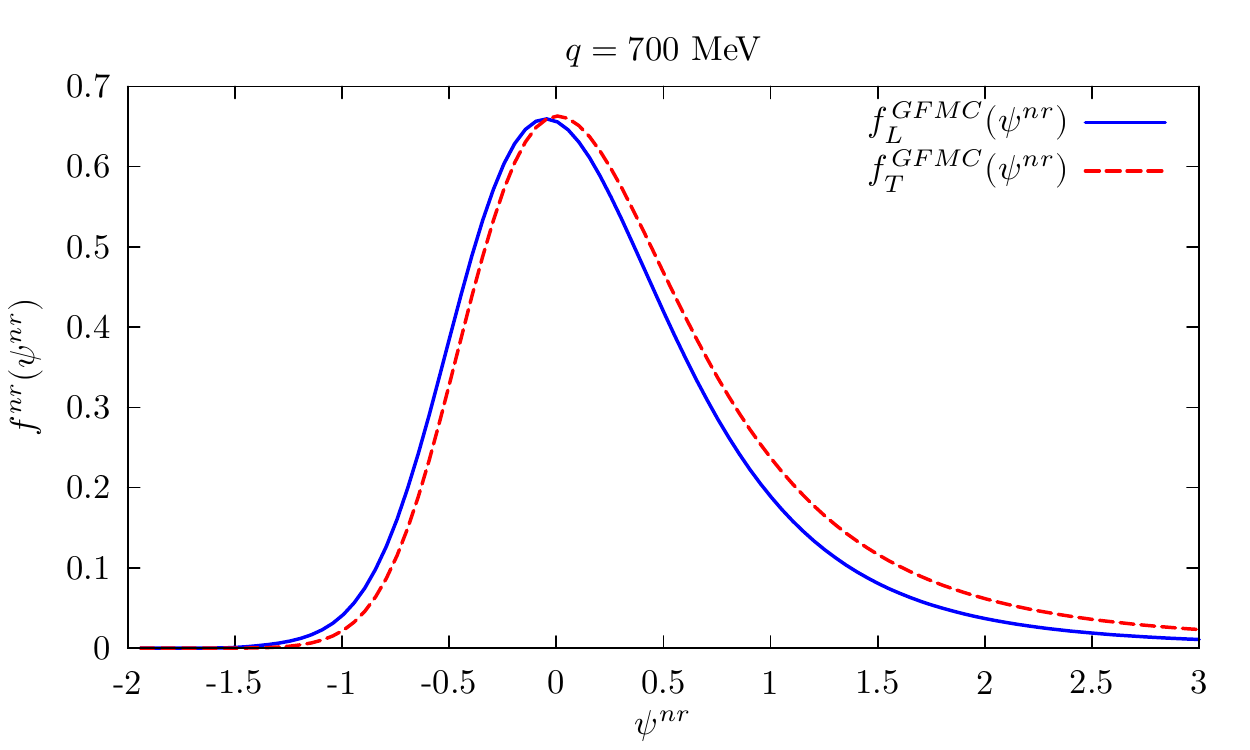}
\includegraphics[scale=0.675]{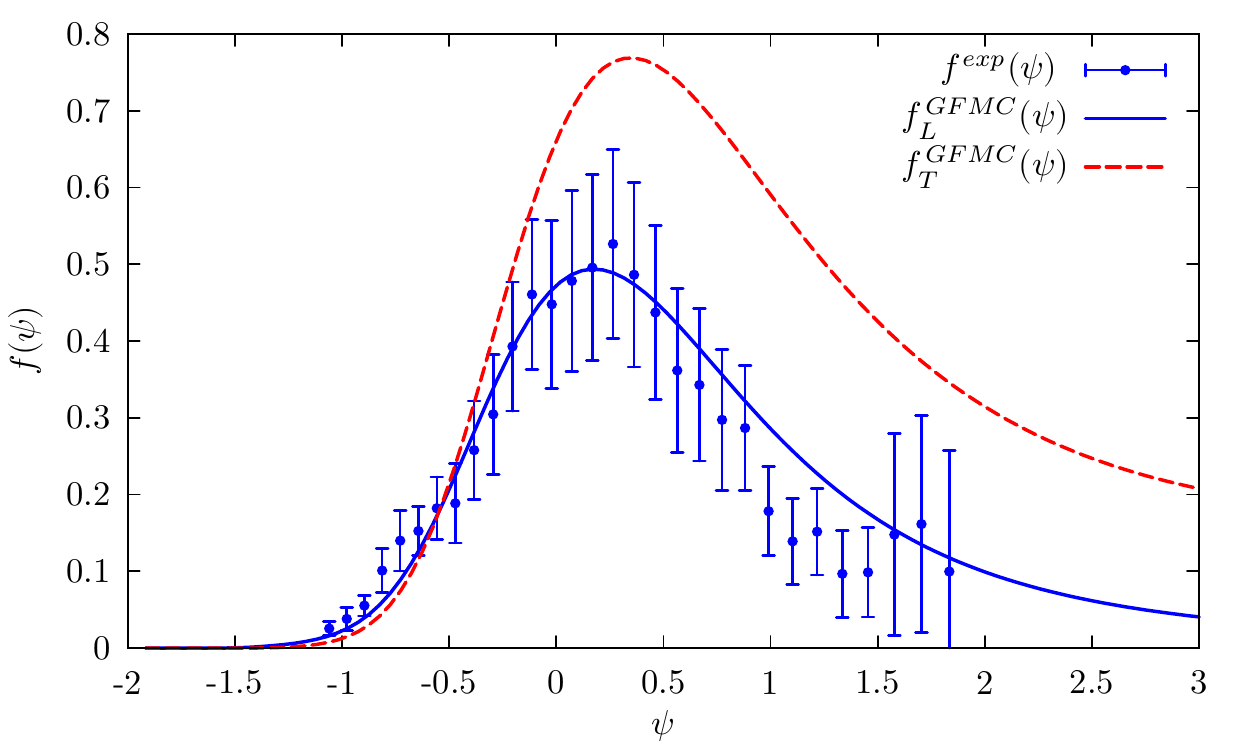}
\caption{ Same as in Fig. \ref{300_4He}  but for  $|{\bf q}|= 700$ MeV.}
\label{700_4He}
\end{figure}
%%%%%%%%%%%%%%%%%%%%%%%%%%%%%%%%%%%%%%%%%%%%%%%%%%%%%%%%%%

%%%%%%%%%%%%%%%%%%%%%%%%%%%%%%%%%%%%%%%%%%%%%%%%%%%%%%%%%%
\begin{figure}[]
\includegraphics[scale=0.675]{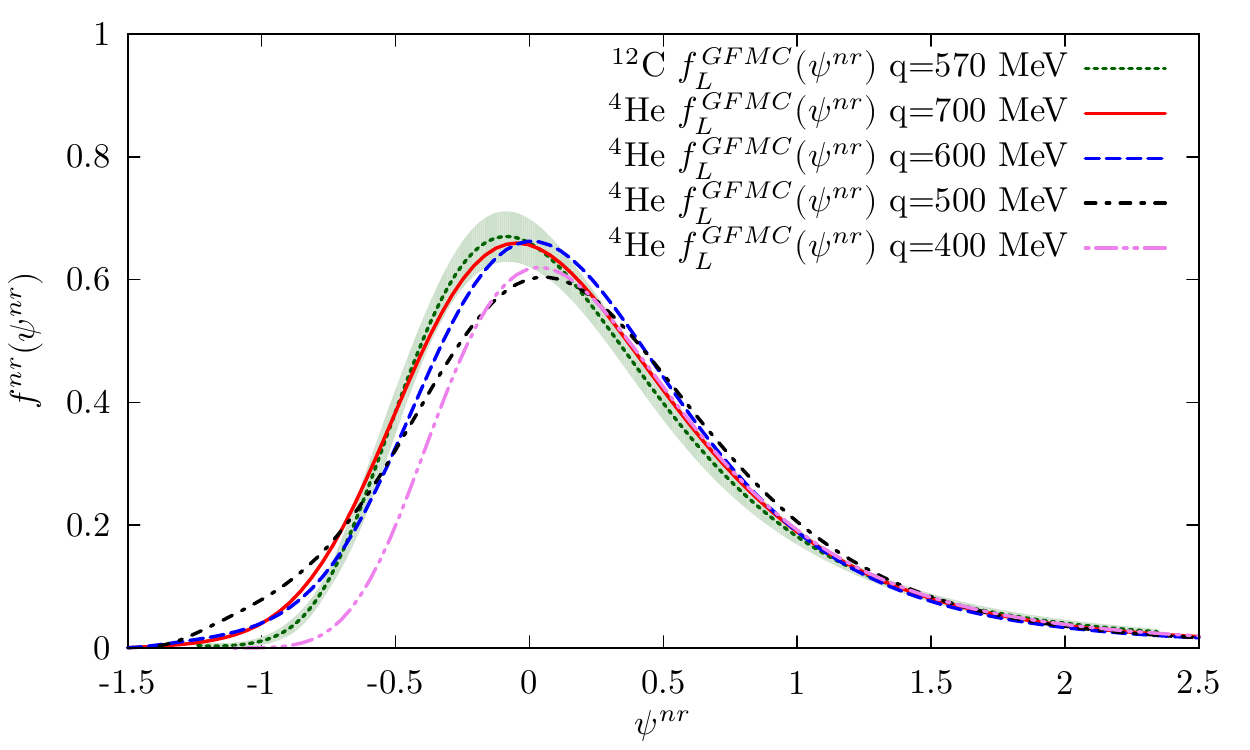}
\vspace*{-.1in}
\caption{(color online) Longitudinal scaling functions obtained from GFMC calculations of the longitudinal response of $^{4}$He for $|{\bf q}|=400,\ 500,\ 600,\ 700$ MeV and of $^{12}$C at $|{\bf q}|=570$ MeV. }
\label{fl_all_4He}
\end{figure}
%%%%%%%%%%%%%%%%%%%%%%%%%%%%%%%%%%%%%%%%%%%%%%%%%%%%%%%%%%

%%%%%%%%%%%%%%%%%%%%%%%%%%%%%%%%%%%%%%%%%%%%%%%%%%%%%%%%%%
\begin{figure}[]
\includegraphics[scale=0.675]{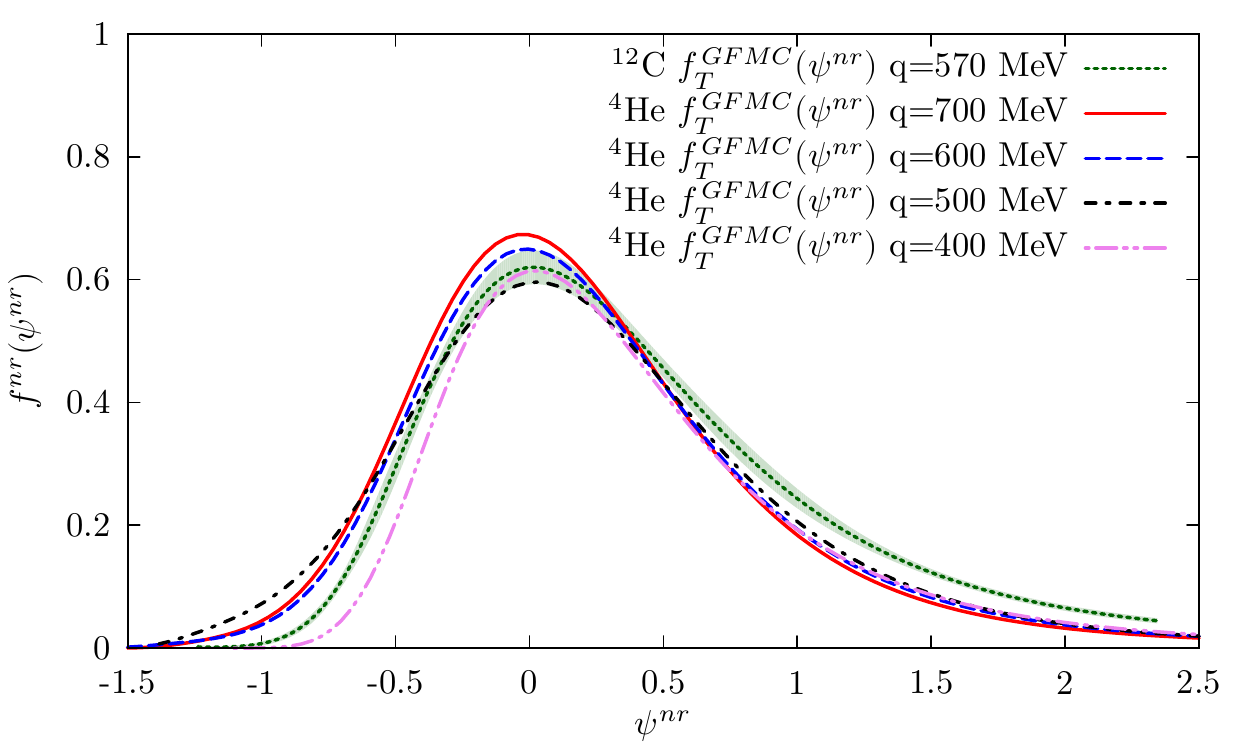}
\vspace*{-.1in}
\caption{(color online) Transverse scaling functions obtained from GFMC calculations of the transverse response of $^{4}$He for $|{\bf q}|=400,\ 500,\ 600,\ 700$ MeV and of $^{12}$C at $|{\bf q}|=570$ MeV.}
\label{ft_all_4He}
\end{figure}
%%%%%%%%%%%%%%%%%%%%%%%%%%%%%%%%%%%%%%%%%%%%%%%%%%%%%%%%%%

In Figs. \ref{300_4He}-\ref{700_4He} we show the longitudinal (solid blue) and 
transverse (dashed red) scaling functions extracted from the GFMC calculations of $^{4}$He at $|{\bf q}|=300,\ 400\ ,500\ ,600,$ and 700 MeV.
In the upper and lower panels the same scheme followed to present the $^{12}$C scaling functions has been adopted. In the longitudinal channel, theoretical calculations and experimental data reported in the lower panels present are in very nice agreement in all the kinematic setups. Finding this agreement up to $|{\bf q}|= 700$ MeV may appear surprising since the GFMC is a non relativistic approach. This can be understood because all the relativistic corrections coming from both the Dirac-spinors and the currents are kept up to $\mathcal{O}[1/m^2]$. However, this is not the case in the transverse channel where relativistic corrections are subleading and have been neglected.  Moreover, the differences in magnitude of the transverse scaling functions, following the discussion carried out for $^{12}$C, are likely to be ascribed to relativistic effects in the prefactors.

The upper panels of Figs. \ref{300_4He}-\ref{700_4He} clearly show that in the $^4$He case the scaling of the zeroth-kind is manifest when the effects of nuclear dynamics are singled out by using the non relativistic expressions for the prefactors. The absence of low-lying transition contributions makes the scaling of the first kind apparent.

The curves of Figs \ref{fl_all_4He} and \ref{ft_all_4He}, where we compare the longitudinal and transverse scaling functions of $^4$He for different values of the momentum transfer,
present a remarkably good scaling behavior. The $^4$He results for $|{\bf q}|= 600\ ,700$ MeV are almost coincident and in good agreement with the longitudinal scaling function of $^{12}$C computed at $|{\bf q}|= 570$ MeV.

Figures \ref{fl_all_4He} and \ref{ft_all_4He} prove that the asymmetric shape of the scaling function does not depend upon the momentum transfer. Consequently, it is not likely to be ascribed to collective excitation modes, that can be accounted for within the random phase approximation.

This analysis, carried out for a variety of kinematics suggests that scaling occurs in the GFMC calculations of the longitudinal and transverse response functions of both $^4$He and $^{12}$C nuclei. Comparing the definition of the longitudinal response function and the one of the corresponding prefactor, see Eq. \eqref{resp:GFMC} and \eqref{g_nr}, while neglecting the spin-orbit contribution, one is lead to conclude that the scaling function corresponds to 
\begin{align}
f_L= \frac{2\kappa\ R_\varrho}{\mathcal{N}}
\label{n:dens}
\end{align}
where $R$ is the nucleon-density response function defined as
\begin{align}
R_\varrho\equiv &\sum_f \langle  0| \varrho^\dagger(\mathbf{q}) | f \rangle  \langle  f | \varrho(\mathbf{q}) |   0 \rangle\, \delta (E_0+\omega-E_f)   \ ,
\end{align}
in terms of the nucleon-density operator 
\begin{equation}
\varrho\equiv \sum_i e^{i \mathbf{q}\cdot \mathbf{r}_i} \frac{(1\pm\tau_{i,z})}{2}\, ,
\end{equation} 
where the $\pm$ applies to protons and neutrons, respectively. Note that Eq.\eqref{n:dens} holds also in the relativistic case, provided that relativistic expressions for the energies are used and  spinors are normalized as $\bar{u}u= \sqrt{m/E}$ to absorb the factor $m^2/(E({\bf p})E({\bf p+q}))$ of Eq. (\ref{eq:scale_integrand}).

On the other hand, the transverse scaling function corresponds to the spin-response, which reduces to the nucleon-density response defined above in the limit of high momentum-transfer, where the impulse approximation is expected to be accurate and where $|\mathbf{q}| \gg |{\bf p}_T|$.
We found that the longitudinal and transverse response functions obtained retaining only the one body current operator scale to the same universal scaling function: the nucleon-density response function.

The results presented in Ref. \cite{Lovato:2016gkq} show that two-body currents lead to a significant enhancement of the transverse response of $^{12}$C in the region of the quasi elastic peak. We expect that the inclusion of this contribution in the scaling analysis, while leaving the longitudinal scaling function unchanged, would contribute to the observed scaling violation of the experimental scaling function in the transverse channel for $\psi\geq0$.

\section{Conclusions}
\label{conclusion}
We have performed a scaling analysis of the GFMC electromagnetic response functions of $^4$He and $^{12}$C for a variety of kinematic setups. Despite the non relativistic nature of the calculation, all the GFMC scaling functions analyzed are strongly asymmetric, with a tail extending to the large $\psi$ region. Within the present picture, this is a consequence of nuclear correlations in both the initial and final states. This is at variance with the findings of Ref.~\cite{Caballero:2007tz}, where the asymmetric shape was ascribed to  relativistic effects in the treatment of the final state interactions. In this regard, it is interesting to point out that the symmetry of the scaling function is not recovered even for momentum transfer as low as $|{\bf q}| =300$ MeV in both $^4$He and $^{12}$C.
When the nuclear dynamics is properly singled out, the $^{12}$C response function shows a fairly good scaling behavior. However, the presence of the low lying transitions which is known to affect the longitudinal channel, introduces non trivial difficulties in drawing definitive conclusions. A better understanding is given by the analysis of $^4$He responses which are free from the uncertainties coming from these contributions. Our results for this nucleus indicate that both the zeroth- and first-kind of scaling occur. Moreover, the $^4$He and $^{12}$C scaling functions fulfill scaling of the second kind once the Fermi momentum of $^4$He is appropriately tuned. 

From our analysis, a novel interpretation of the scaling function emerges. If the spin-orbit contribution to the density-current operator is neglected, it can be easily noted that the longitudinal scaling function corresponds to the nucleon-density response. In the transverse channel, for sufficiently large momentum transfer the term proportional to the
transverse momentum of the incoming nucleon can be safely neglected and the scaling function is proportional to the spin-response. In nuclei characterized by total spin $S=0$, such as $^4$He, $^{12}$C, $^{16}$O and $^{40}$Ca, in the impulse approximation the spin-response reduces to the nucleon-density response. Our findings on the occurrence of zeroth-kind scaling are consistent with this interpretation. 
In fact, within GFMC the scaling violation of the transverse response in the quasi-elastic region is likely to come from two-body currents. This was first noted by the authors of Ref.~\cite{Carlson:2001mp} in which a better agreement between the experimental data and the theoretical calculation of the Euclidean responses of $^3$He and $^4$He was found, once that this term was accounted for.

The role played by two-body current contributions in the electromagnetic responses of $^{12}$C have been recently investigated in Ref.~\cite{Lovato:2016gkq}, where a significant enhancement of the transverse response is observed at all momentum transfers: not only in the {\it dip} region, but in the whole quasi-elastic peak region, extending below the pion-production threshold. In the pioneering work of Ref. \cite{Fabrocini:1996bu}, it has been shown that such enhancement is mainly due to the interference between one- and two-body currents leading to single knock out final state.
In this case the kinematics would be very similar to the those analyzed in this paper, where only one-body current contributes. Hence, we expect that it would be possible to define an appropriate scaling function for these processes. The consequences of the two-body current contribution in the GFMC scaling functions as well as the study of the scaling properties of the total nuclear response \textemdash including both one- and two-body terms\textemdash will be the subject of a future work. 

\section*{Acknowledgements}

Research partially supported by the Spanish Ministerio de Econom\'ia y Competitividad and the European Regional Development Fund, under contracts FIS2014-51948-C2-1-P and SEV-2014-0398, by Generalitat Valenciana under contract PROMETEOII/2014/0068, and by the U.S. Department of Energy, Office of Science, Office of Nuclear Physics,
under contract DE-AC02-06CH11357 (A.L.). Under an award of computer time provided by the INCITE program, this research used resources of the Argonne Leadership Computing Facility at Argonne National Laboratory, which is supported by the Office of Science of the U.S. Department of Energy under contract DE-AC02-06CH11357.

\bibliography{biblio}

\end{document}